\documentclass[usenatbib]{mn2e} 
\usepackage{graphicx,subfigure,times}

\newcommand{\be}{\begin{equation}}
\newcommand{\ee}{\end{equation}}
\newcommand{\bea}{\begin{eqnarray}}
\newcommand{\eea}{\end{eqnarray}}
\newcommand{\bc}{\begin{center}}
\newcommand{\ec}{\end{center}}

\title[Quasar feedback in Galaxy groups]{Effects of Quasar Feedback in Galaxy Groups.} 

\author[Bhattacharya et al.]
       {\parbox{18cm}{Suman~Bhattacharya$^{1}$\footnotemark[1], Tiziana~Di~Matteo$^{2}$\footnotemark[2] and Arthur~Kosowsky$^{1}$\footnotemark[3]} \vspace{0.3cm}\\ 
       $^1$Department of Physics and Astronomy, University of
Pittsburgh, Pittsburgh, PA 15260, USA\\ $^2$Department of Physics, Carnegie-Mellon
       University, 5000 Forbes Ave., Pittsburgh, PA 15213, USA} 

\begin{document}

\maketitle
\begin{abstract}

We study the effect of quasar feedback on distributions of baryons and properties of intracluster medium in galaxy groups using 
high-resolution numerical simulations. We use the entropy-conserving Gadget code that includes gas cooling and star formation, modified to include a physically-based model of quasar feedback. For a sample of ten galaxy group-sized dark matter halos with masses in the range of $1$ to $5\times 10^{13} M_{\odot}/h$, star formation is suppressed by more than 50\% in the inner regions due to the additional pressure support by quasar feedback, while gas is driven from the inner region towards the outer region of the halos. As a result, the average gas density is 50\% lower in the inner region and 10\% higher in the outer region in the simulation, compared to a similar simulation with no quasar feedback. Gas pressure is lowered by about 40\% in the inner region and higher in the outer region, while temperature and entropy are enhanced
in the inner region by about 20-40\%. The total group gas fraction in the two simulations generally 
differs by less than 10\%.  We also find a small change of the total thermal Sunyaev-Zeldovich 
distortion, leading to 10\% changes in the microwave angular power spectrum at angular scales below 
two arcminutes. 

\end{abstract}


\footnotetext[1]{sub5@pitt.edu}, \footnotetext[2]{tiziana@lemo.phys.cmu.edu},\footnotetext[3]{kosowsky@pitt.edu}

\section{Introduction}

Galaxy clusters and groups are the largest gravitationally bound objects in the universe, and
they dominate the total baryon content of the universe. Their spatial distribution and mass 
function contain information about the formation and evolution of large-scale structure, which
in turn constrain a variety of fundamental cosmological properties including normalization of the
matter power spectrum, the cosmic baryon density, and dark matter properties. However, in
order to use them as a cosmological probe, it is necessary to understand their astrophysical
properties, and in particular their baryon physics. This issue is of particular current interest
due to upcoming arcminute-resolution microwave sky surveys like ACT \citep{kosowsky06,fowler07}
and SPT \citep{ruhl05}, which will image galaxy clusters via the Sunyaev-Zeldovich distortions
to the cosmic microwave blackbody spectrum from the hot electrons in the cluster gas \citep{sz80}. 


The majority of baryons in clusters and groups are in the form of hot intracluster gas rather than than individual galaxies. Properties of the Intracluster Medium (ICM) have been studied through a combination of X-ray and radio observations \citep{nulsen05, heinz02, fabian2000}. Although the
dark matter distribution in galaxy clusters follow a self-similar relation \citep{pointecouteu05,vikhilin06}, the hot gas does not \citep{sanderson03,popesso05}. Additional non-gravitational sources of heating are required to explain the observations. One interesting and plausible possibility is the energy radiated from quasars or Active Galactic Nuclei (AGN) and deposited into the ICM
\citep{kaiser91,valageassilk99,nath02, evan05, evan06}, which we study in this work.

The best arena in which to study the impact of various feedback mechanisms is galaxy groups. 
Massive clusters with deeper gravitational potential wells are likely to have their global
thermodynamic and morphological properties less affected by feedback. In comparison, galaxy groups have shallower potential wells while still having enough gas to display the effect of feedback on the ICM.
Galaxy groups have recently been observed in X-rays at redshifts as large as $z=0.6$ \citep{willis05}. 
In the optical band, \cite{sdss07} have compiled group catalogs from the SDSS Data Release 5 
catalog. Evidence for heating by a central AGN or radio source in galaxy groups and clusters has been 
the subject of several recent papers \citep{croston05, jetha06,sanderson05}. These observations show excess entropy in cluster cores, which suggests that  some heating process must act to offset cooling.

In recent years, cosmological simulations including dark matter and gas have
been able to follow the evolution of individual galaxy groups and clusters. A
number of studies have investigated the cluster baryon fraction and its
evolution in numerical simulations. Adiabatic simulations that do not include
radiative cooling find cluster baryon fractions around $0.85$ of the universal
baryon fraction
\cite{evrard90,metzler_evrard94,navarro_etal95,lubin_etal96,eke_etal98,frenk_etal99,mohr_etal99,bialek_etal01}. Preheating
the gas reduces the fraction further
\citep{bialek_etal01,borgani_etal02,muanwong_etal02,kay_etal03}. When cooling,
star formation and other feedback processes are included, the baryon fraction
is higher than that obtained from adiabatic simulations
\citep{muanwong_etal02,kay_etal03,valdarnini03,ettori04,nagai07}. This leads
to an ``overcooling'' problem and indicates an additional feedback mechanism.

In the current study, we analyze the effect of quasar feedback on the baryon
distribution and thermodynamics of hot gas in galaxy groups. We also study its
implication for the Sunyaev-Zeldovich angular power spectrum, which receives a
dominant contribution from high-redshift halos. Ref.~\cite{ks02} showed that
the thermal SZ angular power spectrum provides a strong constraint on the
normalization of the matter power spectrum, $\sigma_8$. Upcoming SZ surveys
like ACT or SPT will have sufficient sensitivity to determine $\sigma_8$ with
an accuracy limited by uncertainty in the theoretical model. Also, the
kinematic SZ effect is a measure of bulk motions in the universe and may be a
competitive probe for studying cosmology \citep{sehgal04, bk06, penn06,
dedeo05, maturi07, roncarelli07}. But one of the major sources of uncertainty
in modeling the kSZ effect is the gas fraction and its evolution. So
understanding both the thermal and kinematic SZ signals requires detailed
understanding of feedback mechanisms in galaxy clusters and groups.  The
mechanisms and effects of feedback are also a long-standing question in
astrophysics, with particular bearing on the process of galaxy formation.

To this end, we have analyzed a sample of ten galaxy groups at $z=1$ from
numerical cosmological simulations of gas and dark matter which have been
extended to include a self-consistent model for the evolution of massive black
holes and their baryon feedback.  At redshift $z>1$, the quasar mode of black
hole accretion is expected to be the dominant feedback mechanism, compared to
the radio-loud accretion mode which becomes important at lower redshifts
\citep{sijacki07}. The size of our simulations prevents studying feedback in
galaxy clusters, but rather restricts us to less massive galaxy groups. But as
already mentioned, galaxy groups with shallow potential wells provide the best
place to study non-gravitational heating and its implications for the
properties of hot gas. High-redshift galaxy groups are also a major
contributor to the thermal SZ power spectrum, which peaks around $z\approx 1$,
when galaxy groups are more numerous than massive clusters \cite{ks02}.

Following this introduction, Section II describes our simulation and its
implementation of quasar feedback. In Section III we study the effect of
numerical resolution on our results; in Section IV we describe our results and
compare them with a simulation that do not include quasar feedback. Finally,
in Section V we summarize our results and discuss directions for future work,
including motivations and prospects for studying more massive galaxy clusters
and more realistic feedback modeling for quasars and AGN.

\section{Simulation}

The cosmological simulations used in this study are described in detail in
\cite{tiziana07}.  They use an LCDM cosmological model with parameters
consistent with the WMAP first-year results \citep{wmap1}: $\Omega_0= 0.3$,
$\Omega_\Lambda= 0.7$, primordial power spectral index $n= 1$, Hubble
parameter $h=0.7$ with $H_0= 100 h \,{\rm km}\,{\rm s}^{-1}\,{\rm Mpc}^{-1}$,
and matter power spectrum normalization $\sigma_8 = 0.9$.  A Gaussian random
initial condition for this cosmology is evolved from high redshifts to the
current epoch using a modified version of the parallel TreePM-SPH code GADGET2
\citep{springel05d}, which manifestly conserve entropy and energy. Gas
dynamics is implemented with the Lagrangian smoothed-particle hydrodynamics
(SPH) technique \citep{monaghan92}. Radiative cooling and heating processes
are computed with a spatially uniform photoionizing UV background
\citep{katz96}. 
For modeling star formation and its associated supernova feedback the
code uses a sub-resolution multiphase model for the interstellar medium developed by
\cite{springel03a}. In this model, a thermal instability is assumed to
operate above a critical density threshold $\rho_{\rm th}$, producing a two
phase medium consisting of cold clouds embedded in a tenuous gas at pressure
equilibrium. Stars form from the cold clouds, and short-lived stars supply an
energy of $10^{51}\,{\rm ergs}$ to the surrounding gas as supernovae. This
energy heats the diffuse phase of the ISM and evaporates cold clouds, thereby
establishing a self-regulation cycle for star formation. The $\rho_{\rm th}$ is
determined self-consistently in the model by requiring that the equation of
state (EOS) is continuous at the onset of star formation. The cloud
evaporation process and the cooling function of the gas then determine the
temperatures and the mass fractions of the two hot and cold phases of the
ISM, such that the EOS of the model can be directly computed as a function of
density. The latter is encapsulating the self-regulated nature of star
formation owing to supernovae feedback in a simple model for a multiphase ISM.
As in the \cite{springel03a} model we have included a model
for supernova-driven galactic winds with an initial wind speed of $v \sim 480
{\rm km s^{-1}}$.

\begin{table}
\begin{center}
\label{tab:simul}
\begin{tabular}{ccccccc}
\hline\hline
Run  &  Box size & $N_{p}$ & $m_{\rm DM}$ & $m_{\rm gas}$ & $\epsilon$& 
$z_{\rm end}$ \\
     &  $h^{-1}$Mpc &&  $h^{-1} M_{\odot}$ &$h^{-1} M_{\odot}$ &  $h^{-1}$ kpc &  \\
\hline
  D4 & 33.75 & $2\times 216^3$& $2.75 \times 10^{8}$ & $4.24 \times 10^{7}$  & 6.25 & 0.00 \\
D6  & 33.75 & $2\times 486^3$& $2.75\times 10^{7}$ &  $4.24\times 10^{6}$& 2.73 & 1.00 \\
\hline\\
\end{tabular}
\end{center}
\caption{Numerical parameters of cosmological simulations (D4 \& D6).} 
\label{sims_table}
\end{table}

A unique aspect of the simulations is their inclusion of supermassive black
holes and the resulting energy feedback from mass accretion \cite{tiziana07}.
Black holes are represented as collisionless ``sink'' particles which grows
from a seed black hole through accretion of mass from its immediately
surrounding gas or through merger with another black hole. Seed black holes of
mass $M=10^5 h^{-1}M_{\odot}$ are placed into the centers of halos whenever
they reach a mass threshold of $10^{10} h^{-1}M_{\odot}$. The subsequent gas
accretion rate onto the black hole is estimated using the
Bondi-Hoyle-Lyttleton parametrization \citep{bondi52,bondi44,hoyle39}.  We
assume a fixed value $\eta=0.1$ for the radiative efficiency $\eta\equiv
L_r/(\dot M_{BH}c^2)$, where $L_r$ is the radiated luminosity and $\dot M_{\rm
BH}$ is the mass accretion rate.  This efficiency value is the mean value of a
radiatively efficient accretion disk onto a Schwarzschild black hole
\cite{ss73}. We further assume that a fraction $\epsilon_f$ of $L_r$ couples
to the surrounding gas in the form of feedback energy $E_f$ deposited
isotropically, i.e. $\dot E_f= \epsilon_f L_r$. A fixed value of
$\epsilon_f=0.05$ is adopted here to fit current data on the normalization of
the $M_{\rm BH}-\sigma$ relation between black hole mass and stellar velocity
dispersion \citep{tiziana05}.

We use three different simulation runs, each of box size $33.75\,{\rm
Mpc}/h$. The box size is a compromise between the requirements of sufficient
spatial resolution to resolve physical processes in high-density regions
surrounding black holes and sufficient volume to allow formation of halos with
galaxy group masses.  We study halos at $z=1$: below this redshift, the
fundamental modes in the cosmological box become nonlinear and the simulations
become unreliable on scales of their largest objects \citep{tiziana03}. We name
 the runs D4 (with and without black holes) and D6 (include black holes)
following the naming scheme adopted in \cite{springel03b}. Runs D4 and D6
include black hole accretion along with cooling, star formation and supernova
feedback, while the run-D4 (no black holes) leaves out black holes but includes
all other physical processes. We use D4 (no black holes) as a baseline
comparison simulation to analyze the effects of quasar feedback on galaxy
groups for the run D4. We also compare D4 and D6 to understand the issues of
resolution and convergence. 
The numerical parameters of the runs, including particle number and mass
resolution, are listed in Table~\ref{sims_table}.


\begin{table}
\begin{center}
\begin{tabular}{ccccc}
\hline
Groups & $R_{200m}$ & $R_{500m}$ & $M_{200m}$ & $M_{500m}$ \\
& Mpc/h & Mpc/h & $10^{13} M_{\odot}$/h &  $10^{13} M_{\odot}$/h \\ \hline
0 & 0.80 & 0.56& 4.71 & 3.08\\
1 & 0.77 & 0.57 & 4.40 & 3.10\\
2 & 0.75 & 0.45 & 2.97 & 1.57\\
3 & 0.68 & 0.46 & 2.14 & 1.64\\
4 & 0.65 & 0.41 & 1.89 & 1.21\\
5 & 0.63 & 0.36 & 1.780 & 0.82\\
6 & 0.63 & 0.37 & 1.783 & 0.84\\
7 & 0.60 & 0.36 & 1.47 & 0.80\\
8 & 0.57 & 0.34 & 1.23 & 0.67\\
9 & 0.53 & 0.36 & 1.13 & 0.76\\ \hline
\end{tabular}
\end{center}
\caption{\rm Properties of galaxy groups in the simulations at $z=1$}
\label{groups_table}
\end{table}

\begin{figure*}
  \begin{center}
    \begin{tabular}{cc}
      \resizebox{85mm}{!}{\includegraphics{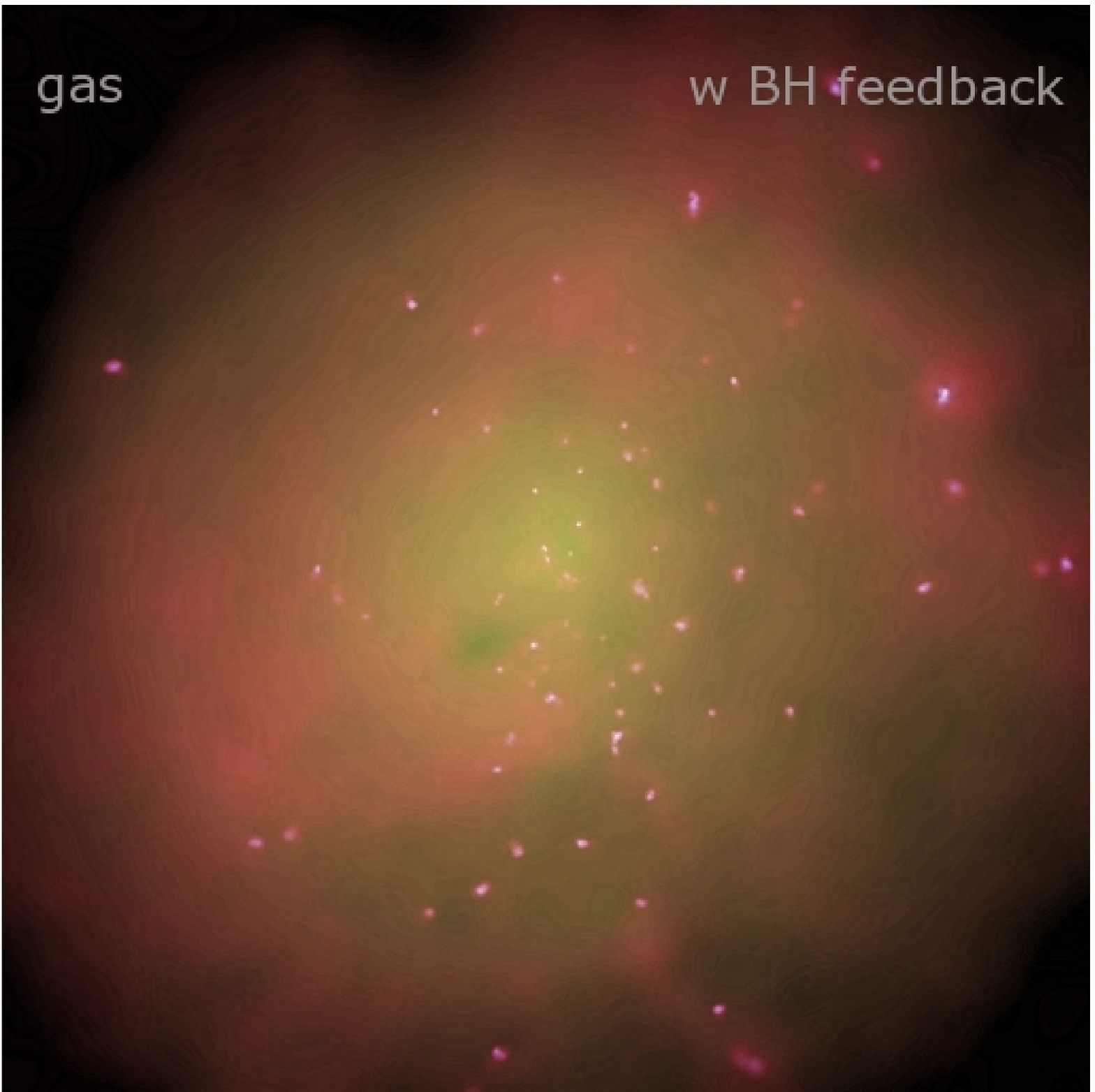}}
       \resizebox{85mm}{!}{\includegraphics{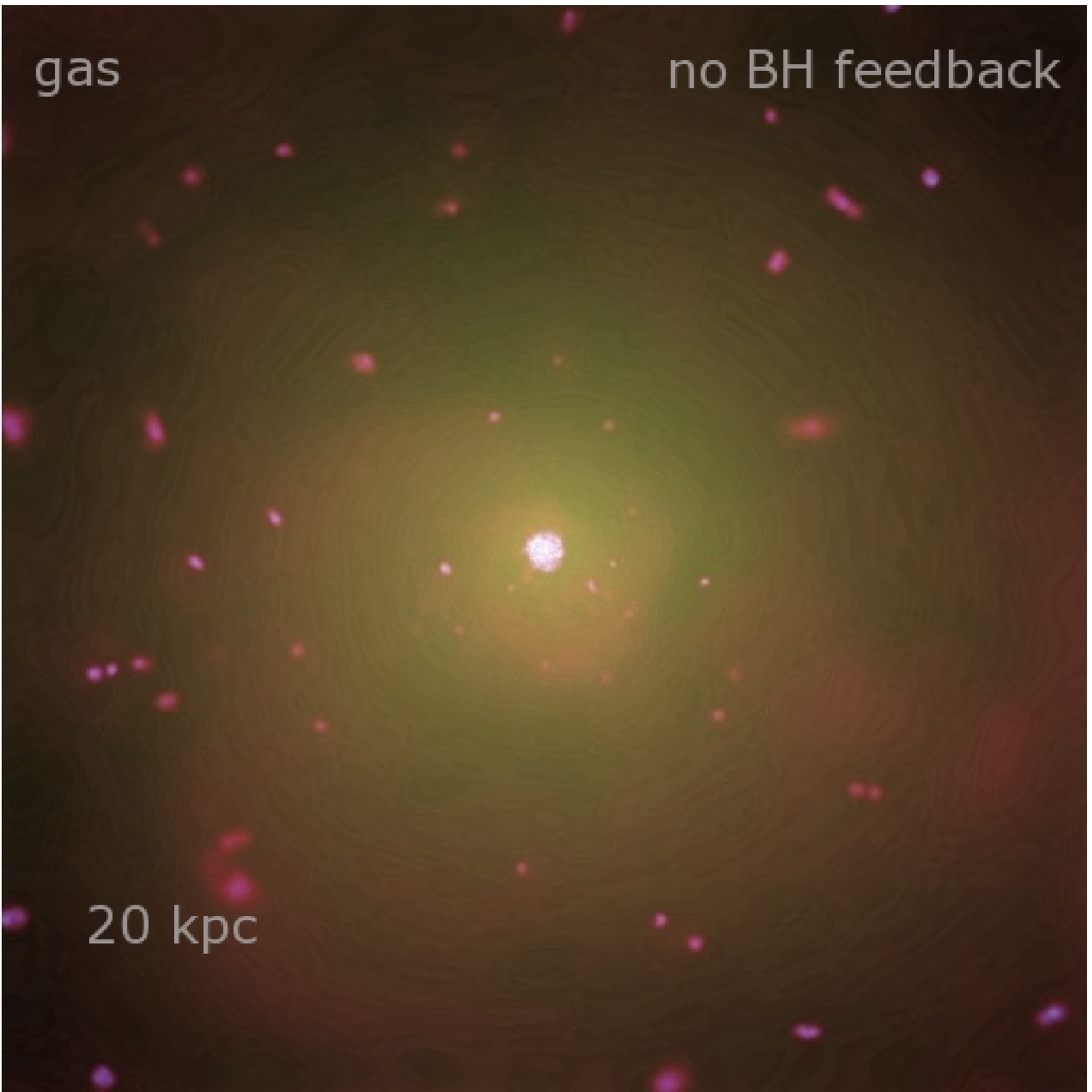}}\\
       \resizebox{85mm}{!}{\includegraphics{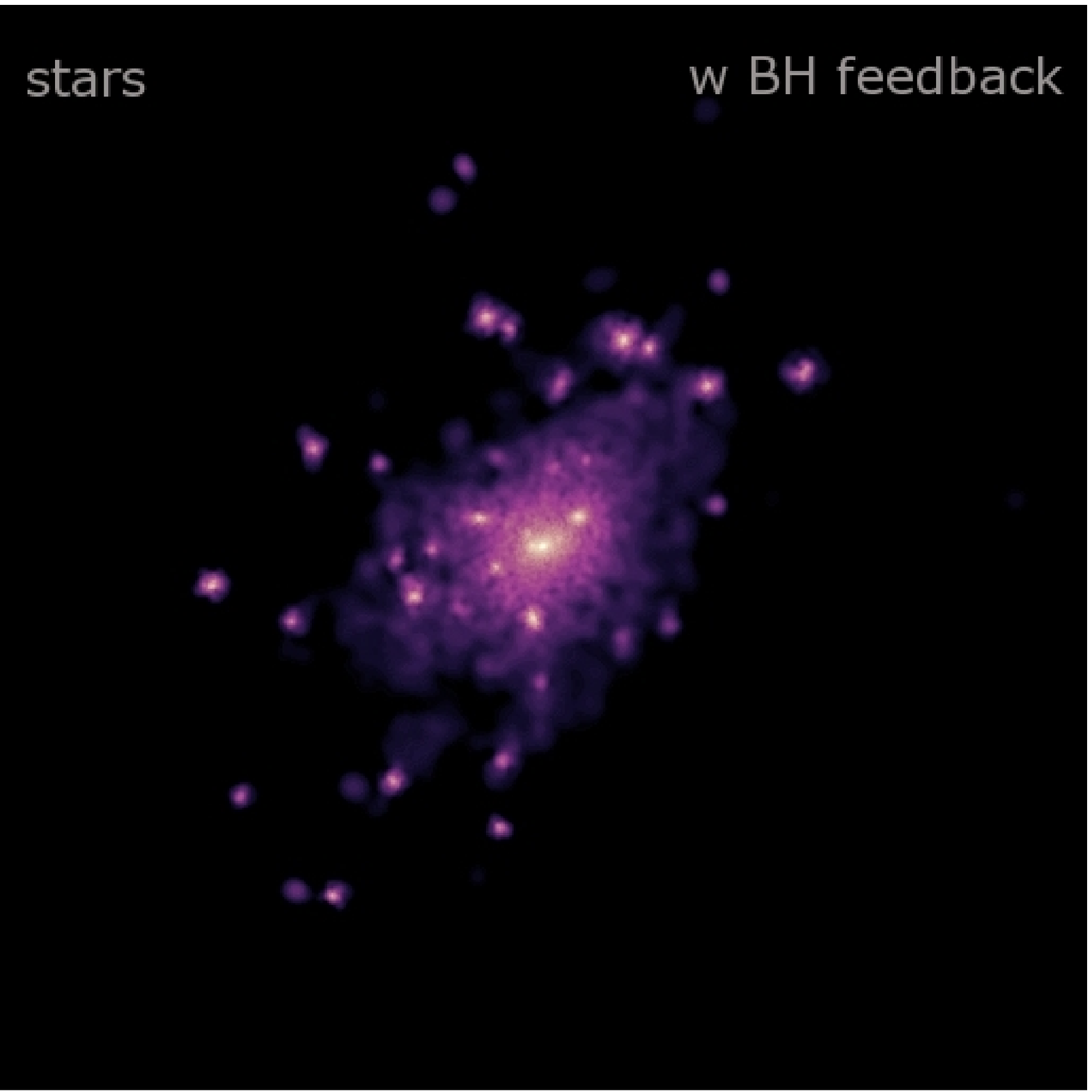}}
       \resizebox{85mm}{!}{\includegraphics{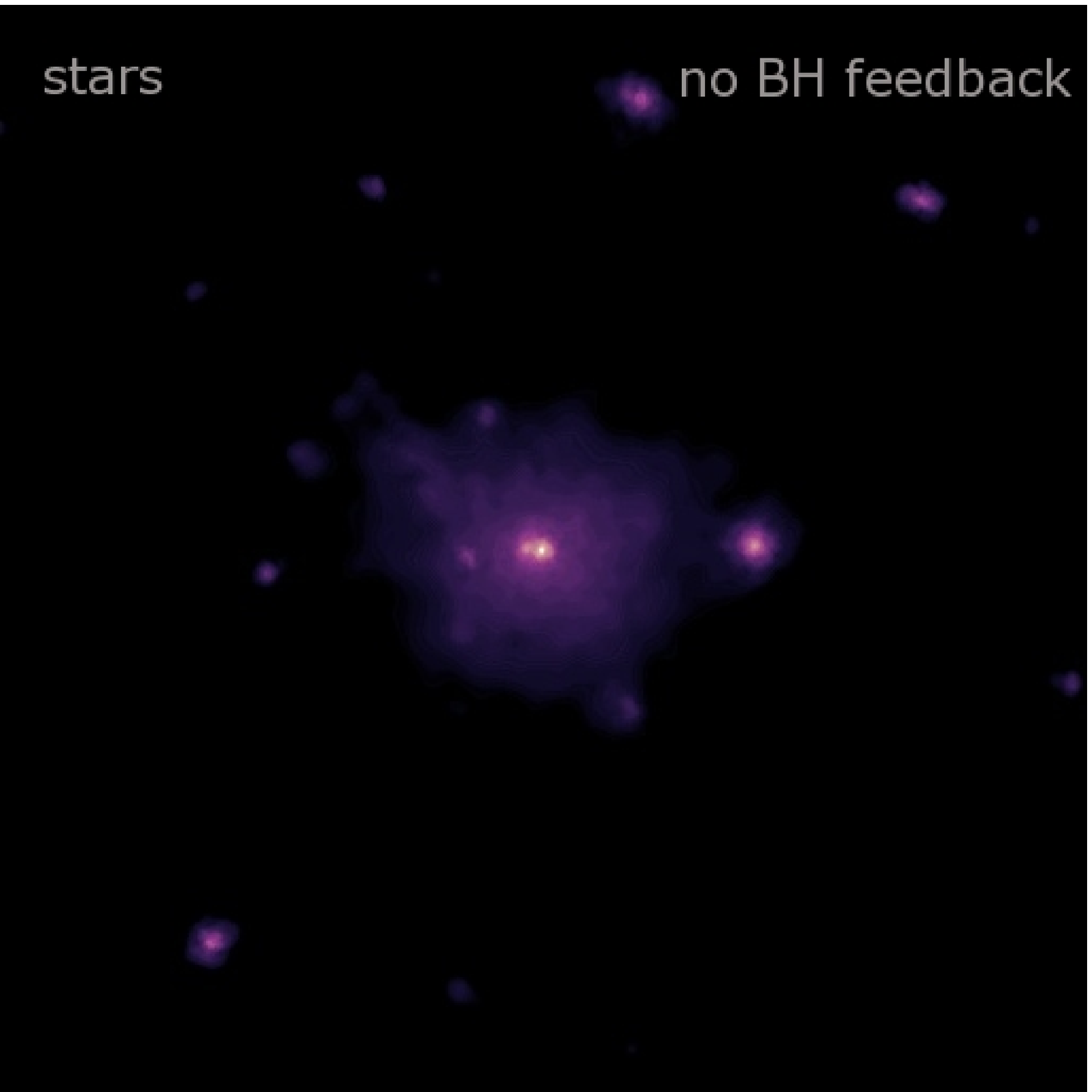}}\\
    \end{tabular}
    \caption{The gas distribution (top) and  star distribution (bottom), both with quasar feedback (left column) and without (right column), for a halo of mass $M=4.6 \times 10^{13} 
M_{\odot}$ at $z=1$. The gas density maps are color coded by temperature
(brightness shows density and color represents temperature). Note the qualitative difference in the distribution of stars between the two simulations. }
    \label{gasmaps}
  \end{center}
\end{figure*}

Table~\ref{groups_table} lists the radius and mass of the galaxy groups formed
in these simulations (the bulk group properties are essentially the same for
all three simulations). Masses are defined as the amount of matter contained
within a spherical region of overdensity 200 ($M_{200m}$) or 500 ($M_{500m}$)
times the mean density of the universe at $z=1$
\citep{tiziana03}. Figure~\ref{gasmaps} shows gas density and star
density for the most massive halo ($M_{200m}=4.7 \times 10^{13} h^{-1}
M_{\odot}$) in the simulation. The left panel shows the map for each of
the properties when black hole feedback is included while the right panel
gives the map with no quasar feedback. Note the gas density maps are color
coded by temperature- the brightness shows the density and the color
represents the temperature. It is evident that the gas is hotter when the
feedback is included compared to when not included. Also the distribution of
stars has changed significantly when quasar feedback is included.


\section {Effects of Numerical Resolution}

To study the effect of quasar feedback, we need to resolve quasars and surrounding gas at kiloparsec scales while simultaneously following the formation and evolution of galaxy groups at megaparsec scales. Given this huge dynamic range, it is worthwhile to check how numerical resolution affects our results. We have run two simulations, namely ``D4" and ``D6," with the same cosmological parameters, initial conditions and simulation volume. The lower-resolution D4 run uses $2 \times 216^3$ total particles, while D6 uses $2 \times 486^3$ particles. The corresponding mass resolution of the gas is $4.24 \times 10^7 M_{\odot}/h$ and $4.24 \times 10^6 M_{\odot}/h$. Their spatial resolution is characterized by gravitational softening lengths of $6.25\,{\rm kpc}/h$ and $2.73\,{\rm kpc}/h$ respectively. 

\begin{figure*}
  \begin{center}
    \begin{tabular}{cc}
       \resizebox{85 mm}{!}{\includegraphics{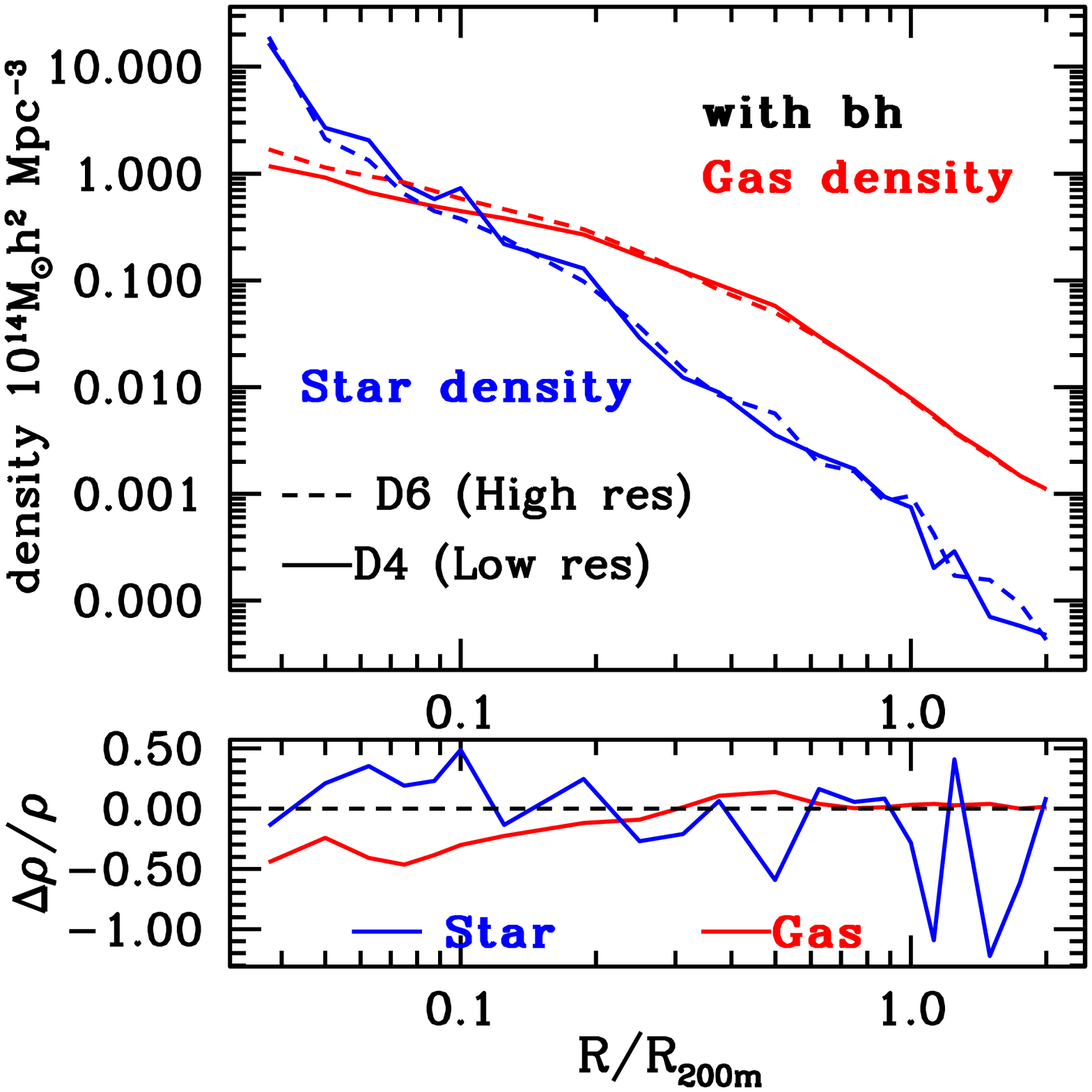}} 
     \resizebox{85mm}{!}{\includegraphics{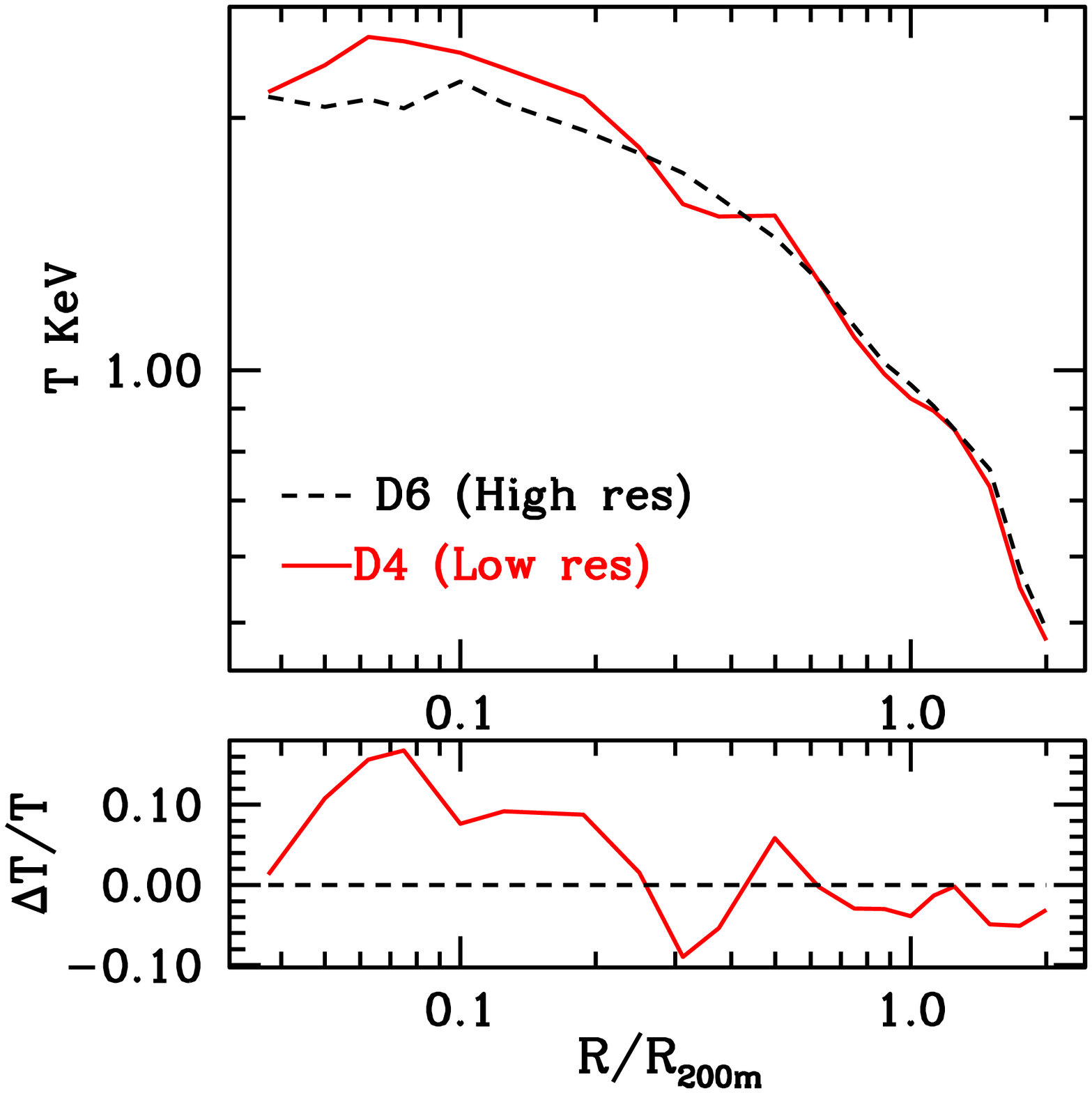}}\\ 
     \resizebox{85mm}{!}{\includegraphics{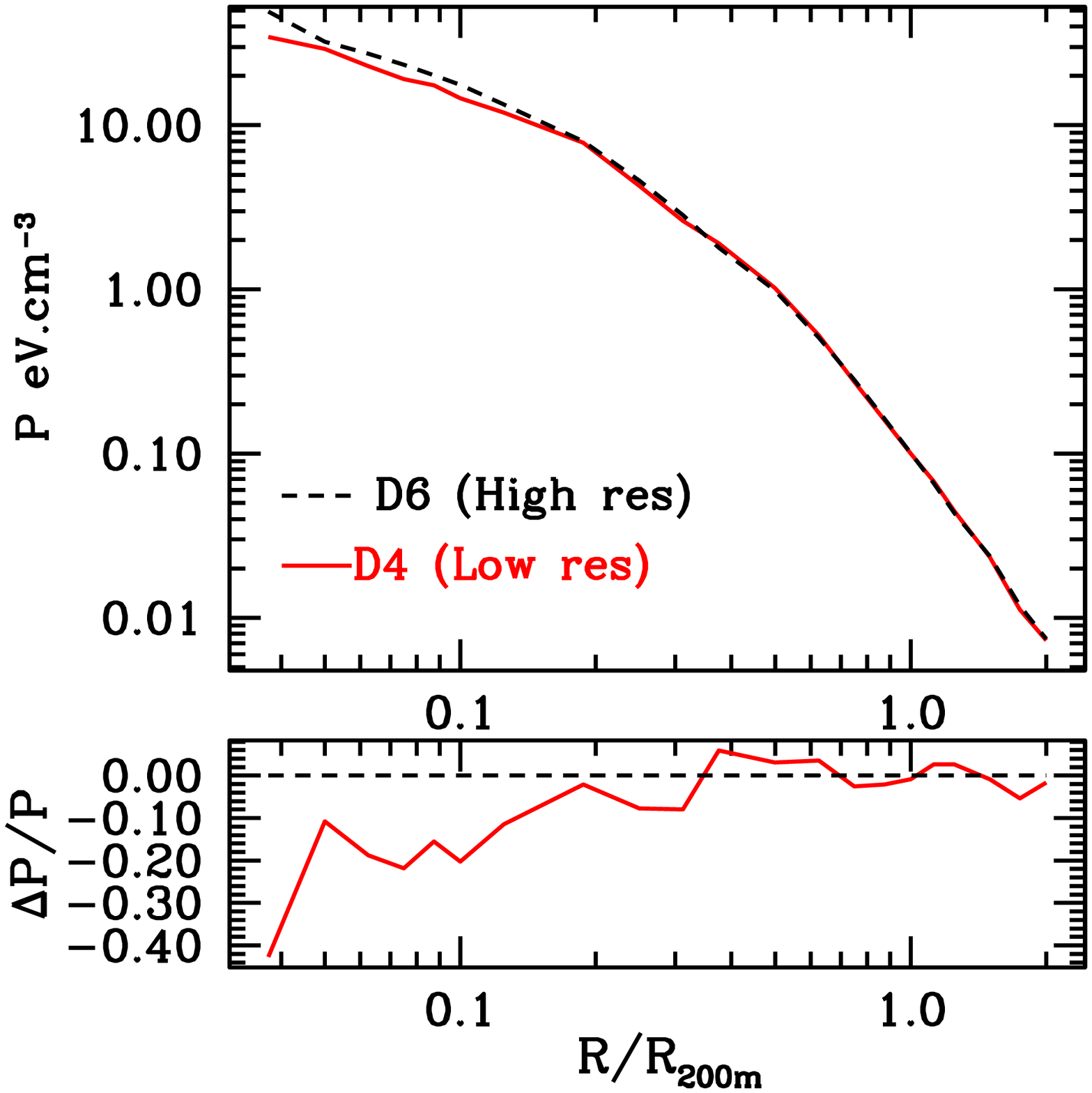}}
      \resizebox{85mm}{!}{\includegraphics{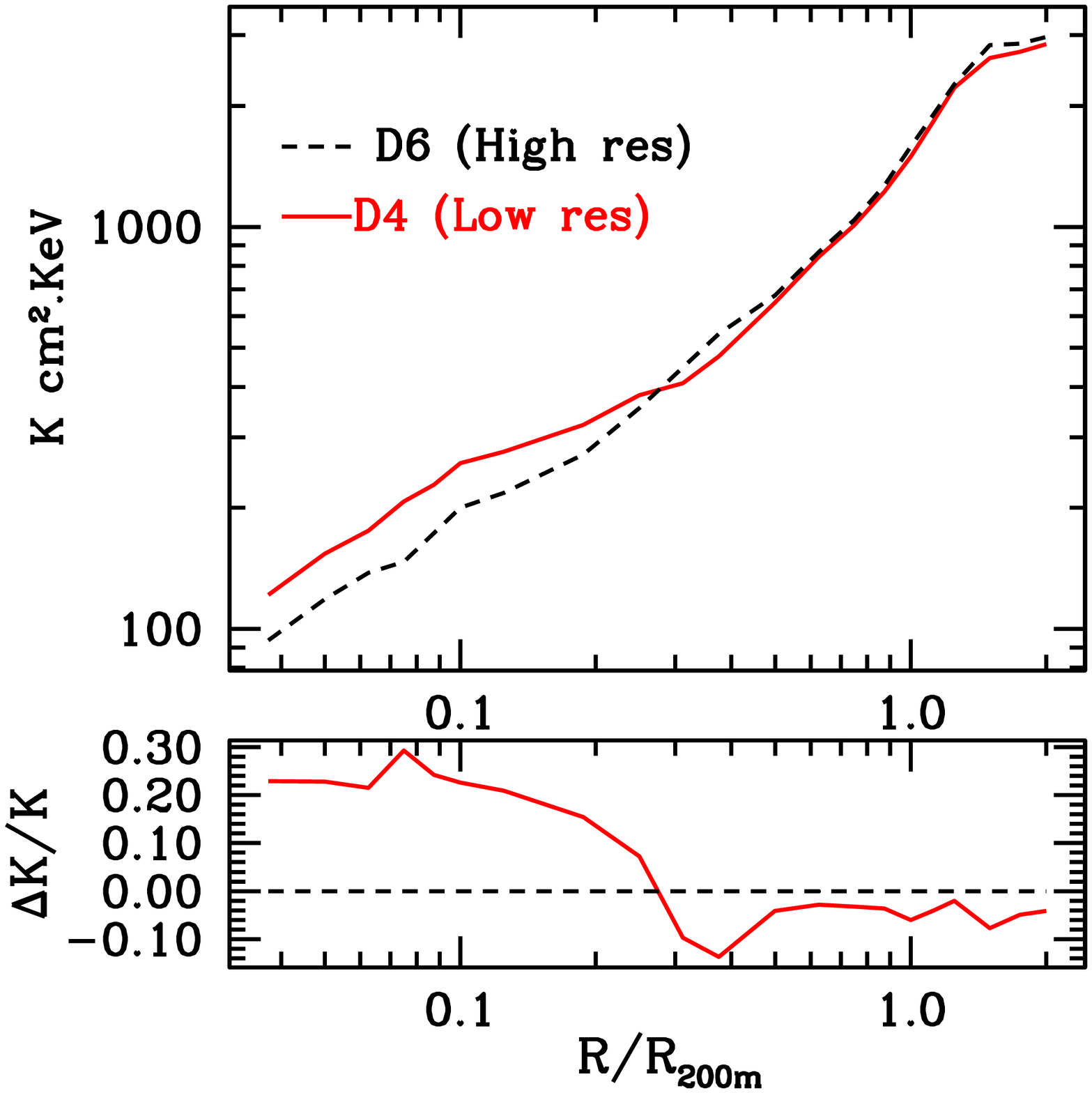}}
\end{tabular}
    \caption{The effect of numerical resolution on various quantities as functions of radius: baryons, both gas and stars (top left ), temperature (top right), pressure (bottom left) and entropy (bottom right). In each panel, dotted lines represents higher resolution (D6) and solid lines represent lower resolution(D4).}
    \label{resolution_fig}
  \end{center}
\end{figure*}

We have studied the difference in the star and gas distributions at redshift $z=1$, with comparisons
displayed in Fig.~\ref{resolution_fig}. These plots show the average differential profile in the simulations.

On average, both star and gas distributions agree within $10\%$ 
for the D4 and D6 runs for $R>0.1 R_{200m}$. Beyond $R=R_{200m}$,
statistical fluctuations causes star distributions to vary. Note that most of the star formation occurs in the inner region of the halo, so these statistical variations in the outer parts do not affect any of the conclusions about star fraction. 

The temperature profile shows roughly $10-15\%$ difference between the simulations D4 and D6 
in the inner region of the cluster, dropping to $5\%$ for $R> 0.2 R_{200m}$. The pressure profile shows relatively more robustness to numerical resolution: a 10\%-15\% difference in the inner region drops to only
$5\%$ to $3\%$ for  $R>0.2 R_{200m}$. Numerical resolution should thus have a minimal effect on 
the thermal SZ flux, since most of the signal comes from outside the core. Finally, the entropy profile shows a difference of 20\% in the inner region and a 5\% difference for $R> 0.3 R_{200m}$.

As already shown in \cite{springel03b, springel03}, using a large number of cosmological simulations, simulation including star formation and cooling converge reasonably well in the resolution range between D4 and D6.



Given this rough quantification of the effect of increased resolution, we proceed to
analyze the lower-resolution D4 simulation in the rest of the paper and compare with the same resolution run without black holes, noting where errors due to numerical resolution limits might be a significant fraction of the effects being discussed. In the following sections, we study the differential and cumulative profile for each physical quantity with two lower resolution D4 runs both with and without including black holes. For each physical quantity we also calculate the difference in the profiles for each halo between the two runs with and without black holes and then show the mean of the difference. Also we find there are atleast 3 mergers namely 2nd, 5th and 8th most massive halos(in the group of 10 halos) we considered here. While studying  the average profiles, we have excluded these halos from the averaging process so that the profiles do not get biased. However we have reported properties of all the halos when they are studied as a function of mass.



\section{Results}

\subsection{Thermodynamics of the Intracluster Medium}

\begin{figure*}
  \begin{center}
    \begin{tabular}{cc}
           \resizebox{85mm}{!}{\includegraphics{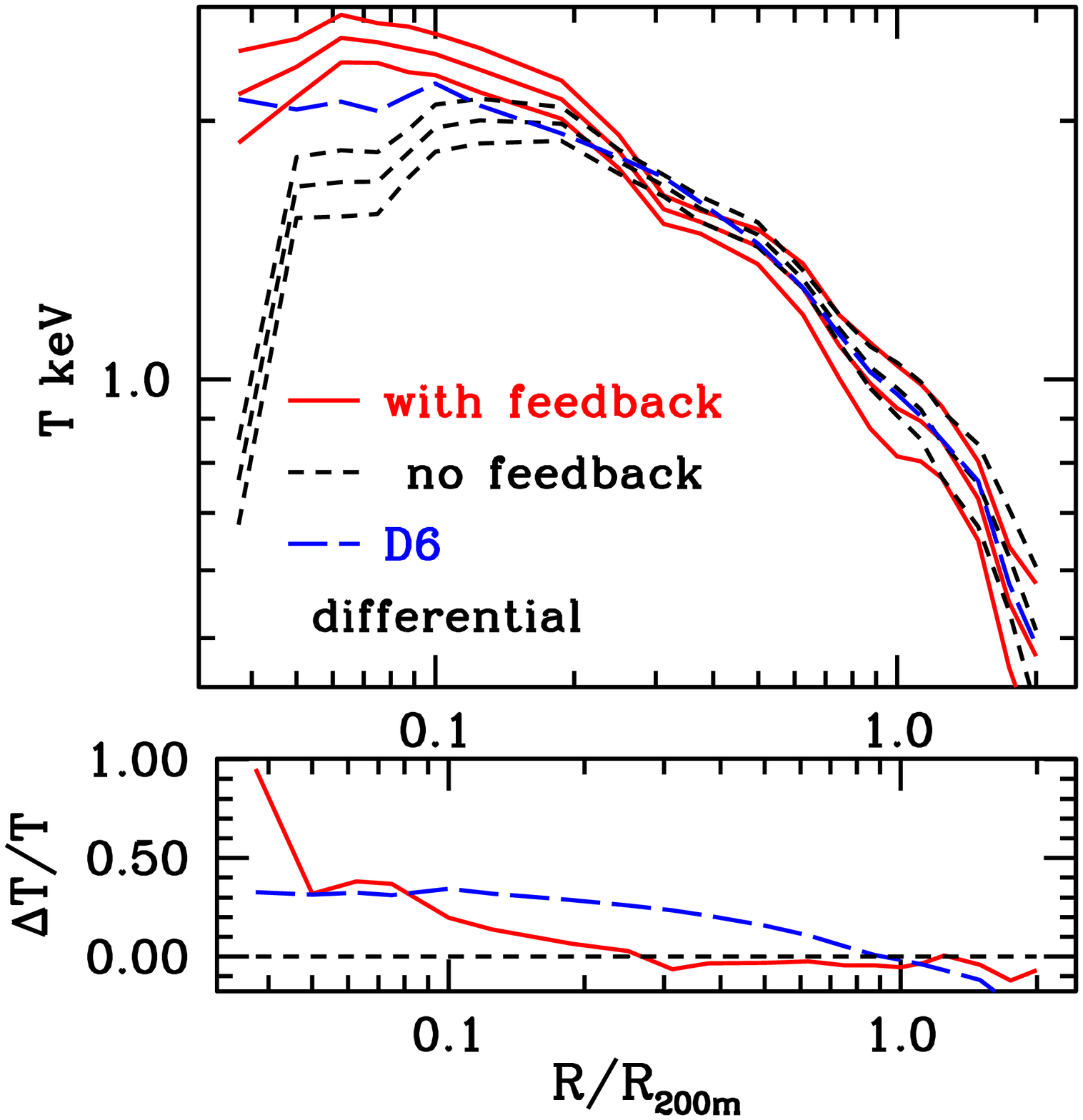}}
       \resizebox{85mm}{!}{\includegraphics{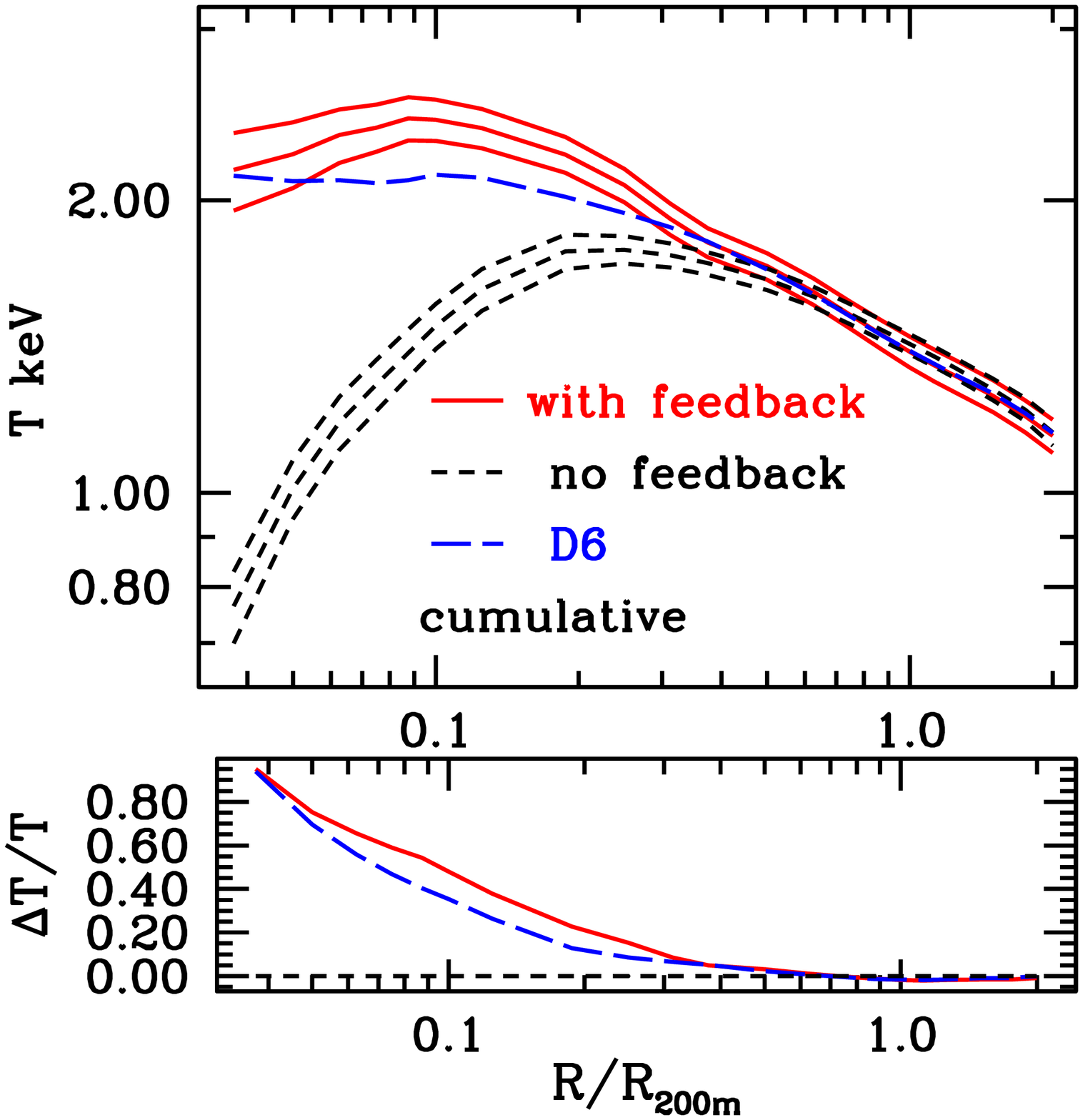}} \\
    \end{tabular}
    \caption{The mean differential (left) and cumulative (right) temperature profile of gas averaged over seven halos. For each top panel, solid lines represent the mean and scatter around the mean profile for simulation D4 including quasar feedback, while the dotted lines represents the same quantities for simulation D4 with no quasar feedback. Also shown is the mean profile from the D6 run (blue dashed line. The lower panels show the mean fractional change between the halos in the two runs. The blue dashed line shows the mean and the scatter in the difference in the profiles between D4 and D6 ( resolution effect) while the solid red line shows similar difference between the D4 runs ( the effect of including the black holes) }
    \label{temp_profile}
  \end{center}
\end{figure*}

In this Section we study the impact of quasar feedback on thermodynamics of the ICM, namely on the three
quantities pressure, temperature and entropy. Figure~\ref{temp_profile} gives the average temperature profile with scatter around the mean. In the inner region ($R<0.2 R_{200m}$) of the halo, the temperature is enhanced by about 15-20\% and by 5-10\% in the region $0.2 R_{200m}<R<0.5 R_{200m}$. This is physically 
reasonable as quasar feedback is coupling part of its radiated thermal energy to the surrounding ICM. We do not see any change in temperature due to quasar feedback at radii outside the halo core. For comparison we also show the average mean profile from the D6 run. 

Note however, that the temperature profile inside the halo core becomes steeper when the feedback is included, whereas the observations at low redshift shows a rather flat profile inside the core. This disagreement might be either due to the inability of the feedback mechanism to explain the observed temperature profile and an improved model is needed or that one needs to include other sources of feedback in the simulations. Observations of group size halos at high redshift will be needed in order to understand whether the temperature profile indeed gets steeper at higher redshift or a better feedback mechanism is required to explain the flatness of the temperature profile.

The temperature of the system agrees fairly well with previous studies made using halos of similar mass \cite{borgani04, khalatyan07, finoguenov01}. For eg., a halo of mass $4.7 \times 10^{13} M_{\odot}h^{-1}$ is expected to have a temperature of around 1 keV at z=0. We find a temperature of 1.5 keV for a similar system at z=1. If a virial scaling relation is assumed this translates to a temperature of about 1 keV at z=0 which is consistent with previous studies. 

\begin{figure*}
  \begin{center}
    \begin{tabular}{cc}
           \resizebox{85mm}{!}{\includegraphics{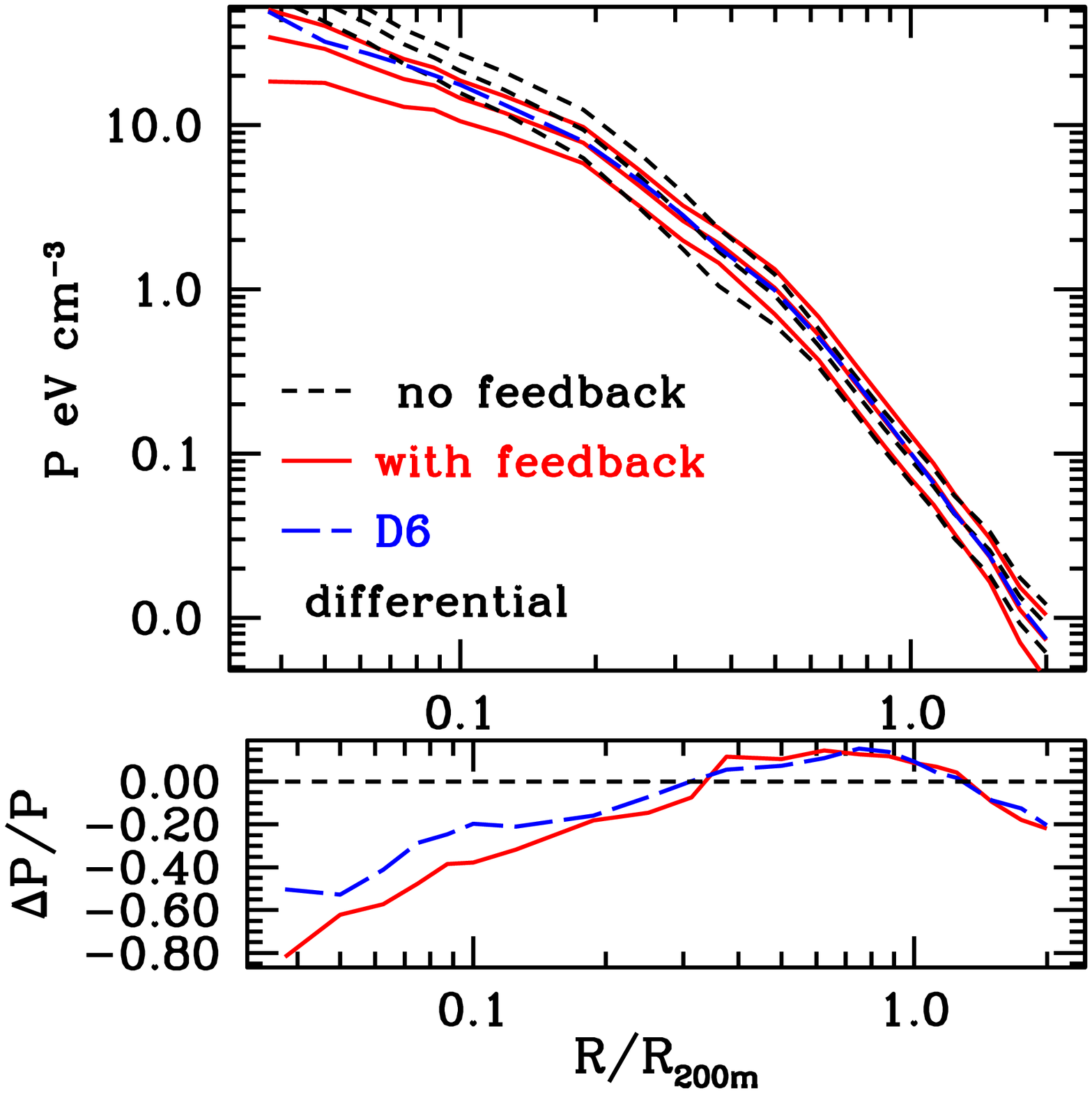}}
       \resizebox{85mm}{!}{\includegraphics{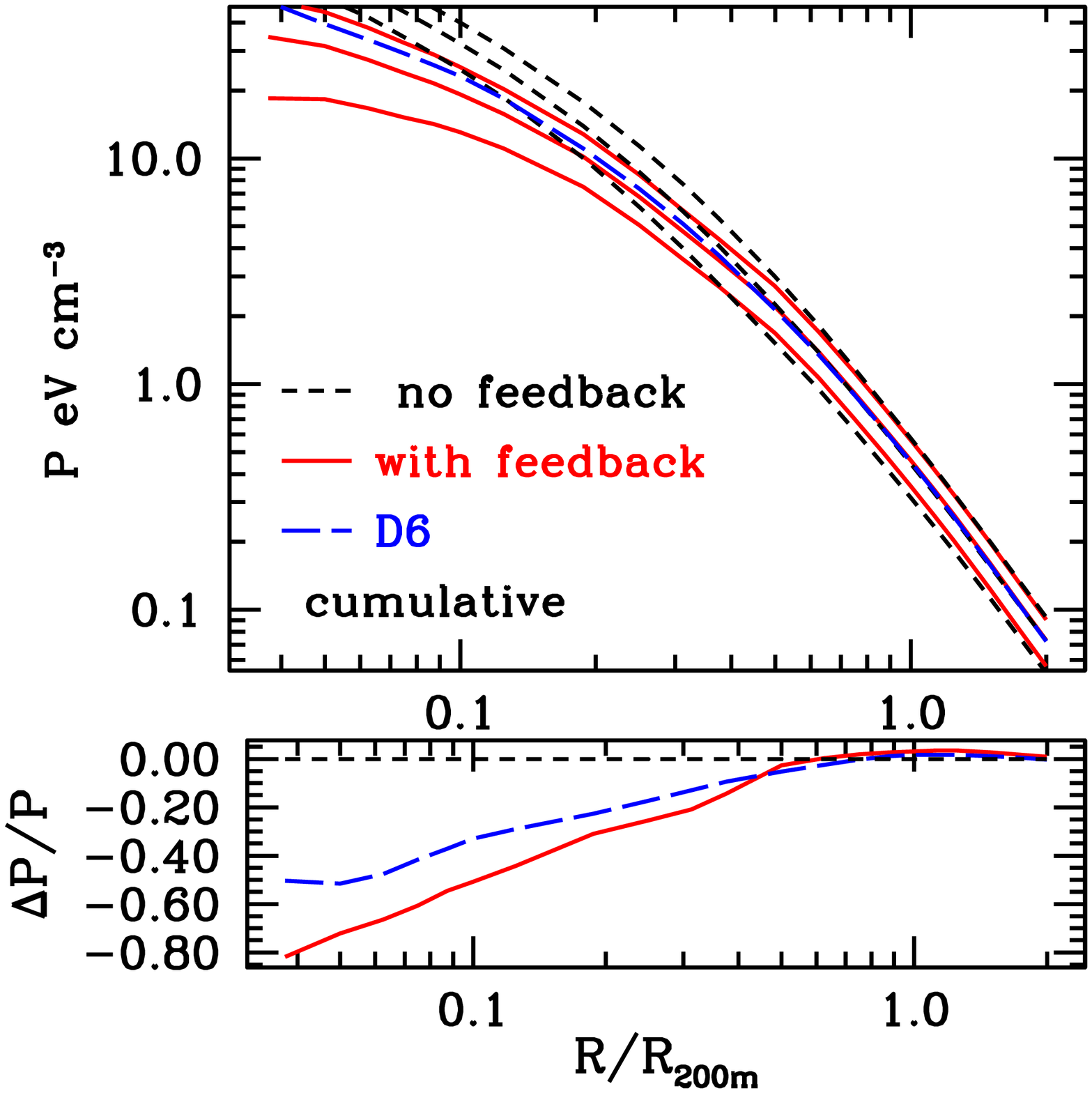}} \\
    \end{tabular}
    \caption{Same as in Fig.~\ref{temp_profile}, except for pressure.}
    \label{pressure_profile}
  \end{center}
\end{figure*}
\begin{figure*}
  \begin{center}
    \begin{tabular}{cc}
           \resizebox{85mm}{!}{\includegraphics{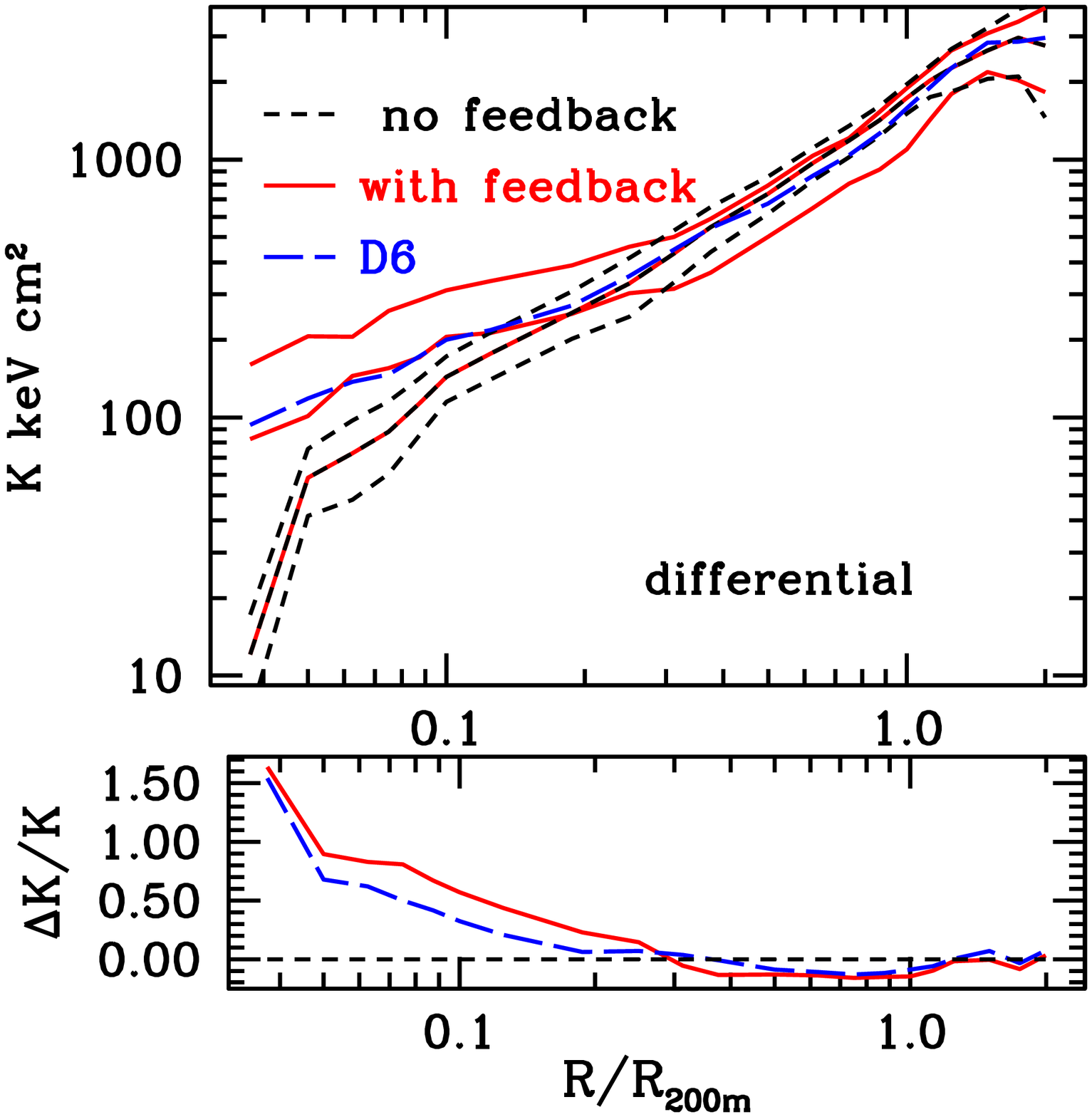}}
       \resizebox{85mm}{!}{\includegraphics{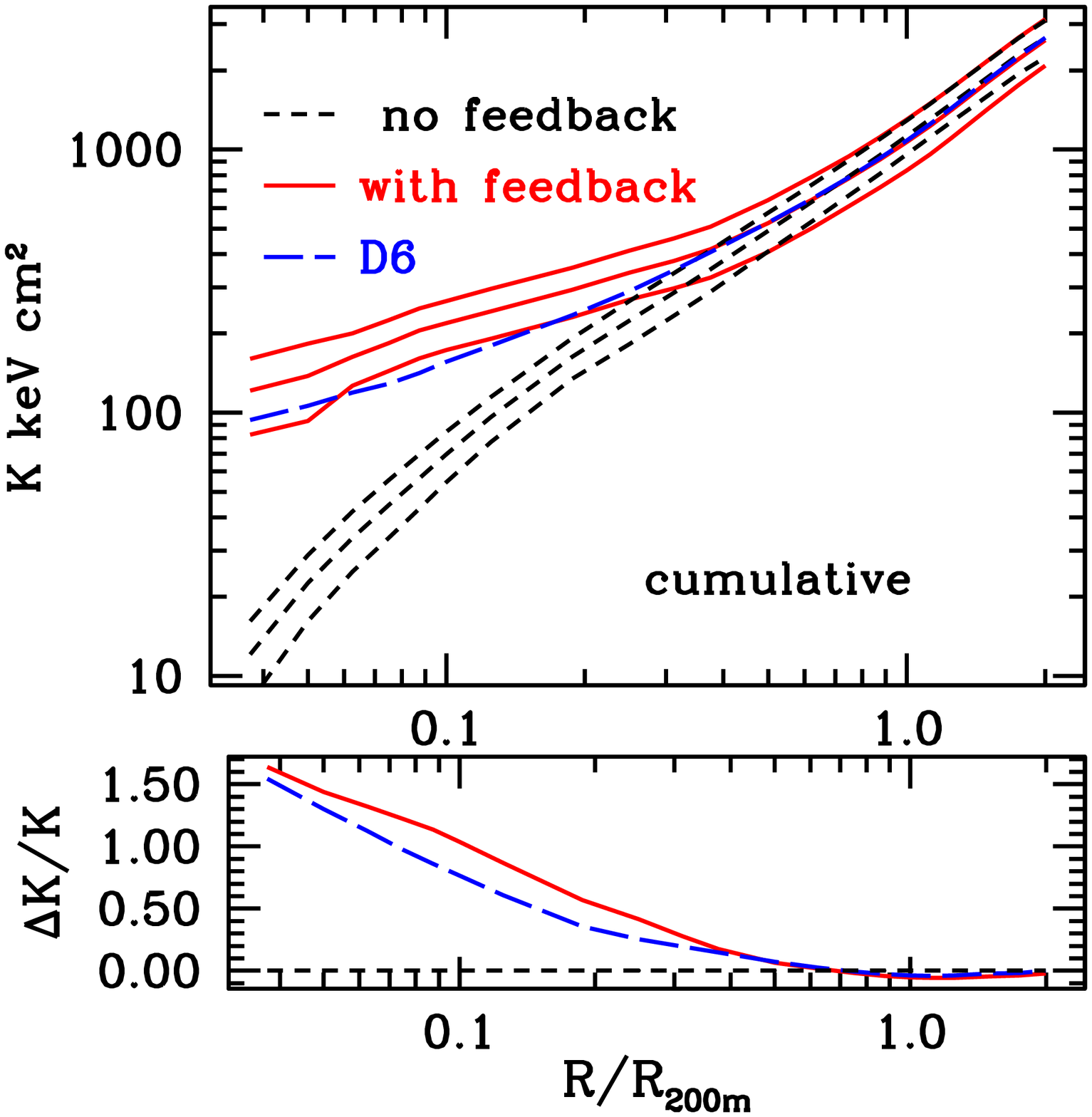}} \\
    \end{tabular}
    \caption{Same as in Fig.~\ref{temp_profile}, except for entropy.}
    \label{entropy_profile}
  \end{center}
\end{figure*}

The corresponding average pressure profile is shown in Fig.~\ref{pressure_profile}. We find that the pressure decreases for $R<0.3\,{\rm Mpc}/h$, beyond which
quasar feedback clearly leads to a pressure enhancement of 15\% to 20\% out to radius of $R_{200m}$. The entropy profile is shown in Fig.~\ref{entropy_profile}. The excess entropy near the core region is 50\% larger than the no feedback case.  The observational finding for the entropy profile for small  groups \cite{ponman03}  agrees fairly well with the current study when virial scaling is assumed to translate the entropy profile at z=1 in the current study to z=0. The scatter around the mean profile for each of these quantities is large, so we need a larger sample size to confirm these systematic deviations. The entropy and pressure profile indicates that the quasar feedback has driven the gas out from the inner region and redistributed in the outer region. The lower panels of the figures show the fractional difference for each quantity. As shown, in the inner region the difference in the profiles is significant;  far in excess of the numerical resolution error. Similar differences can be seen in the outside region where the numerical resolution error is few percent.

\subsection{Baryon Fraction of the Intracluster Medium}

\begin{figure*}
  \begin{center}
    \begin{tabular}{cc}
      \resizebox{85mm}{!}{\includegraphics{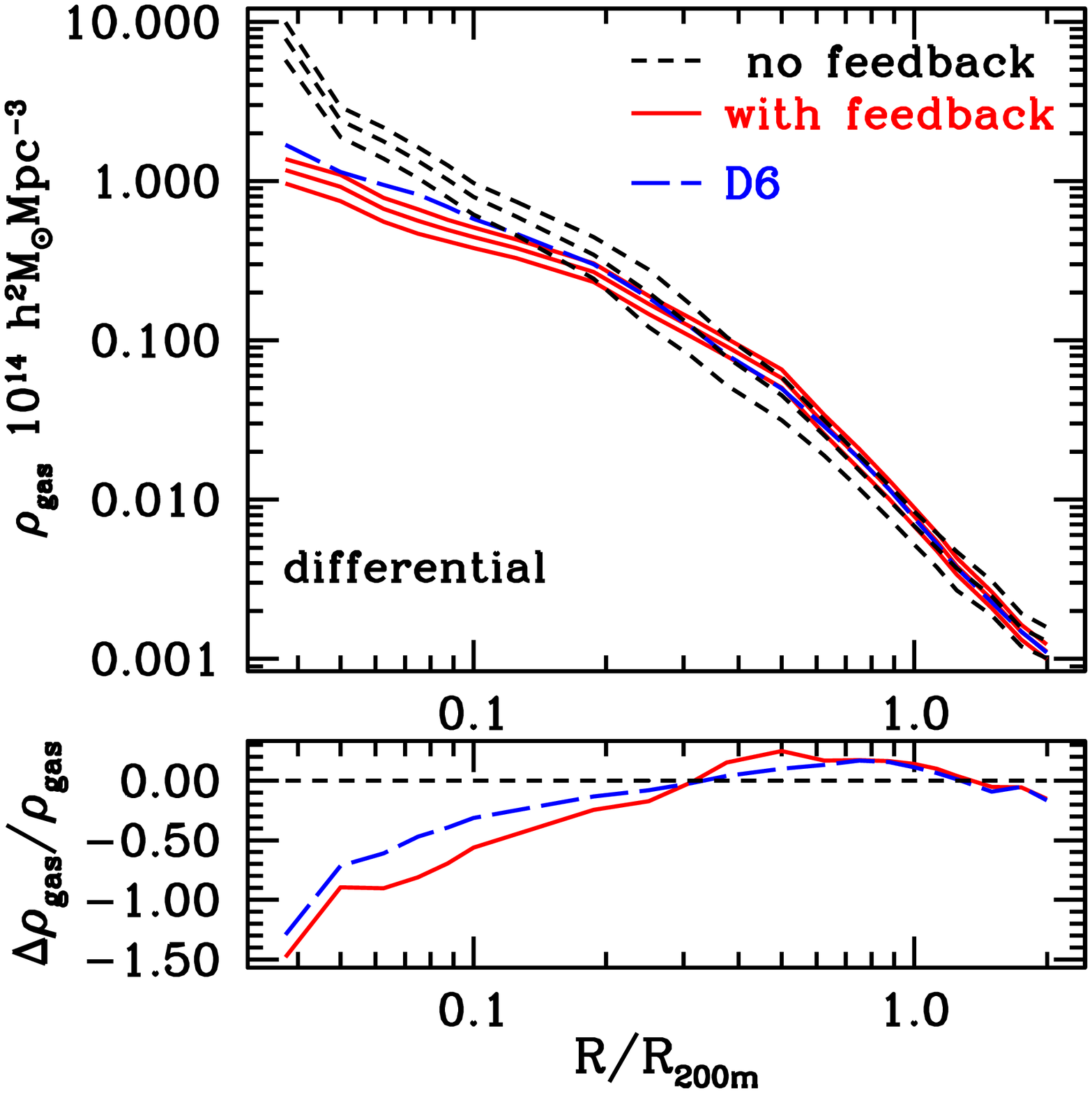}}
       \resizebox{85mm}{!}{\includegraphics{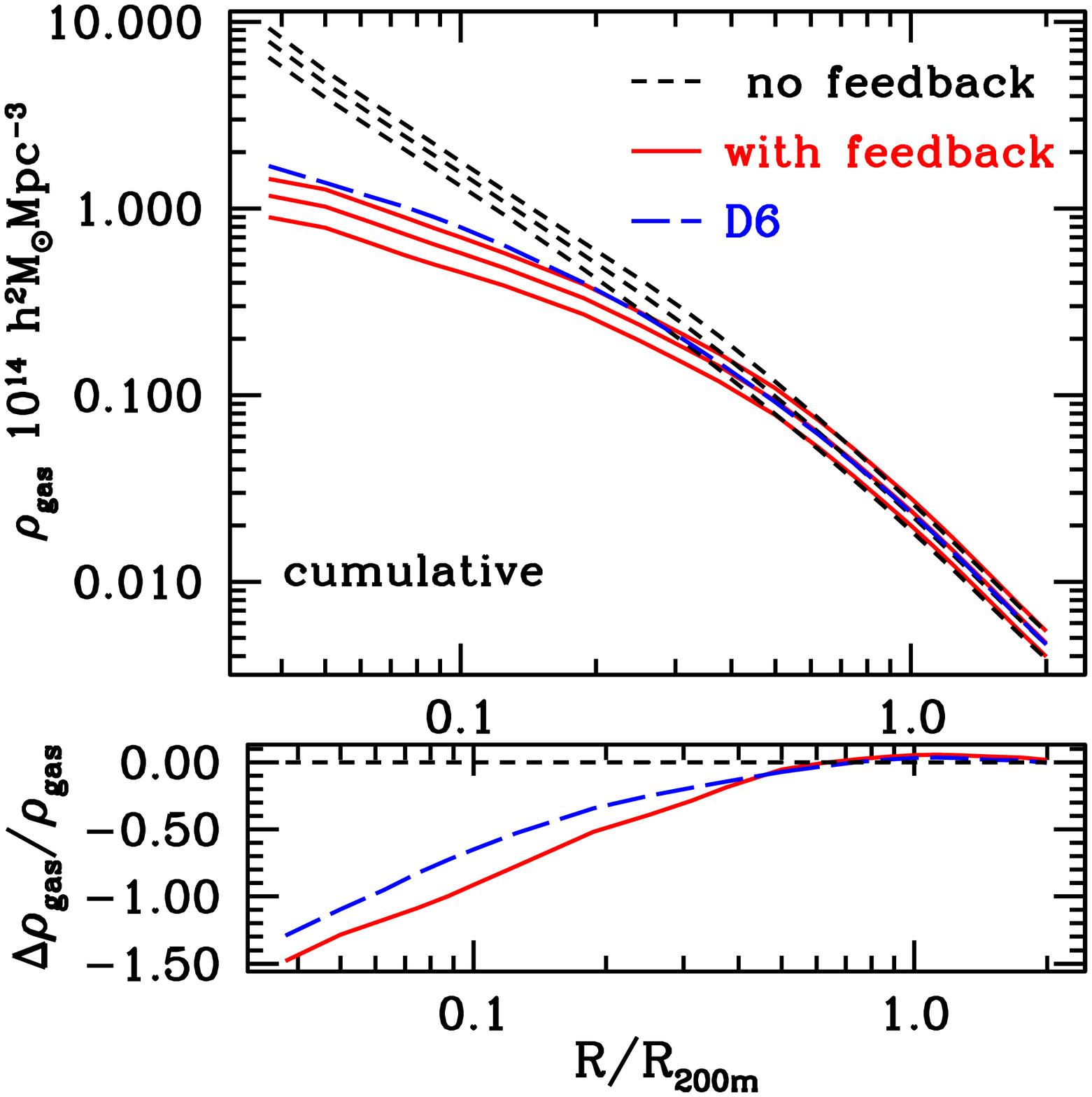}} \\
       \resizebox{85mm}{!}{\includegraphics{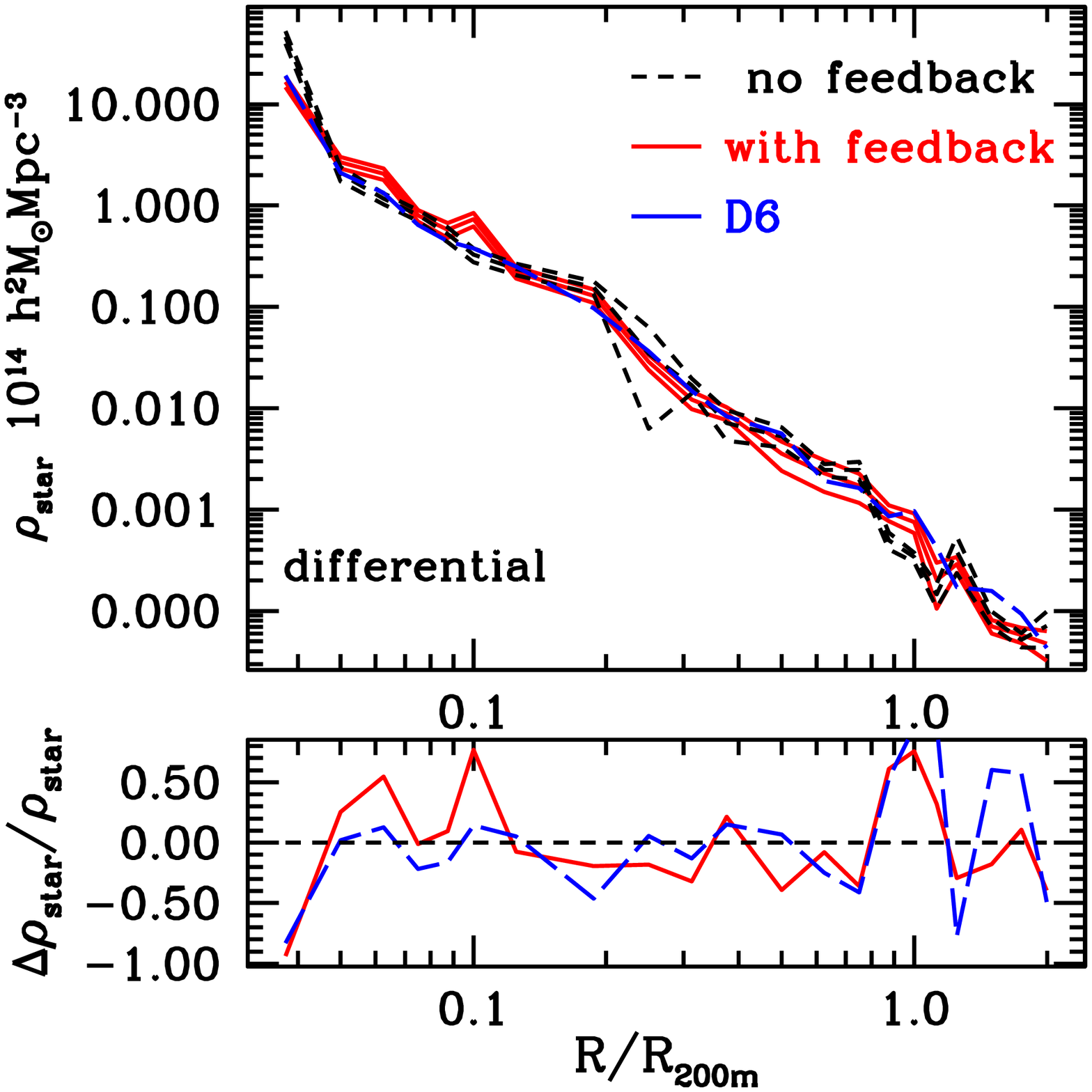}}
       \resizebox{85mm}{!}{\includegraphics{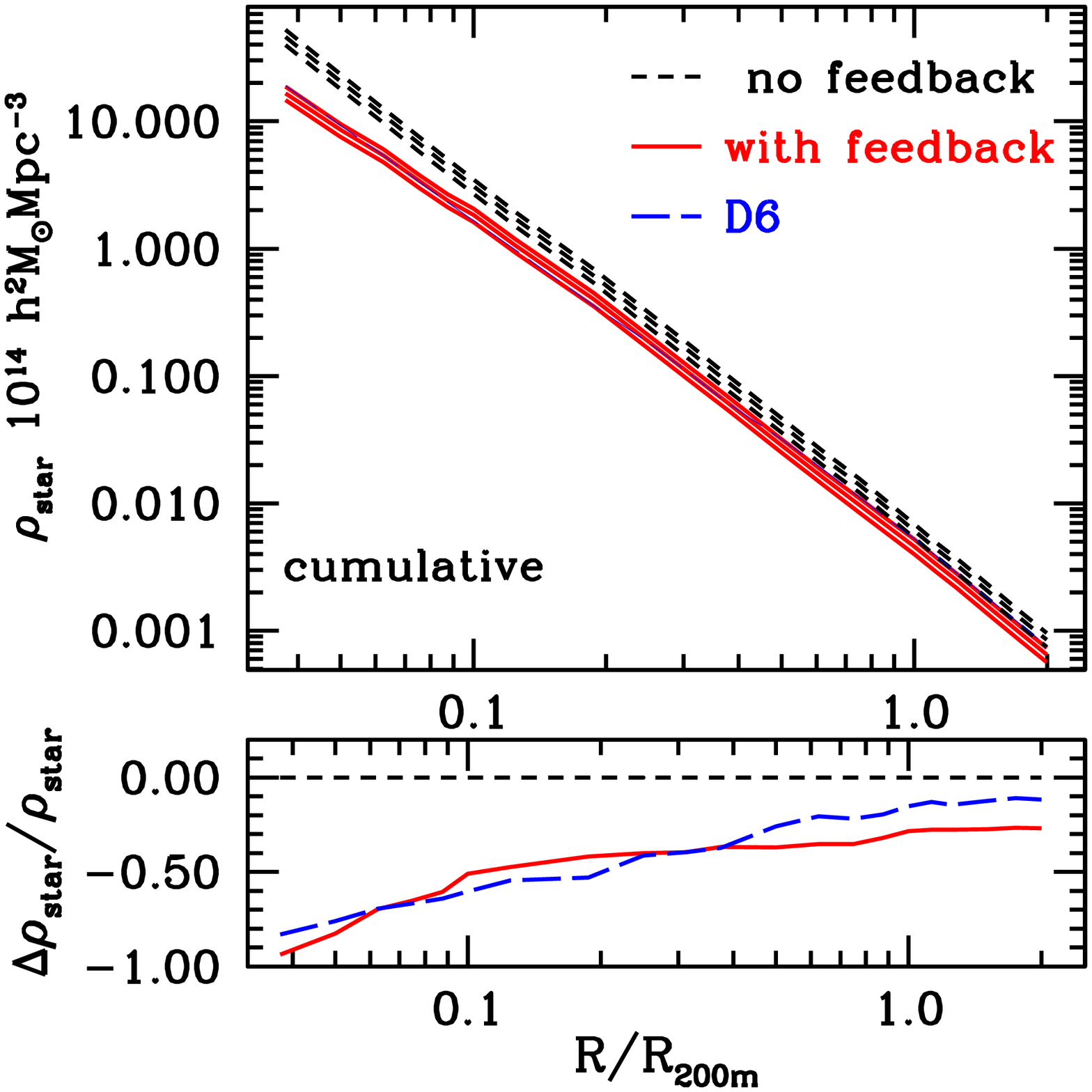}} \\
          \end{tabular}
    \caption{Same as in Fig.~\ref{temp_profile}, except for gas density (top panels) 
    and star density (lower panels).}
    \label{baryon_profile}
  \end{center}
\end{figure*}

A particularly important issue for interpreting future Sunyaev-Zeldovich measurements is the gas
fraction in a given halo. Here
we consider the effect of quasar feedback on both baryonic components, stars and hot gas. The ten most
massive objects formed in the simulation have masses ranging from $1$ to 
$5\times 10^{13} M_{\odot}/h$. Each object is binned in spherical shells, and the mass fractions of stars, gas and dark matter within each shell are
normalized to the primordial baryon fraction $\Omega_b/\Omega_m$. 
Figure~\ref{baryon_profile} shows the average differential (left) and cumulative (right) distribution of gas and stars. Note the difference in star formation between the simulations with and
without quasar feedback is on average 20\% to 40\% out to radius $R=0.6 R_{200m}$. 
It is evident that quasar feedback substantially suppresses star formation at all radii; the cumulative star distribution is 30\% lower when feedback is included. The feedback mechanism provides enough pressure support that a significant amount of gas fails to collapse and form stars. Comparing differential and cumulative profiles, it is evident that most of the star formation is suppressed in the interior region of the halo. 

\begin{figure*}
  \begin{center}
       \resizebox{7 in}{7 in}{\includegraphics{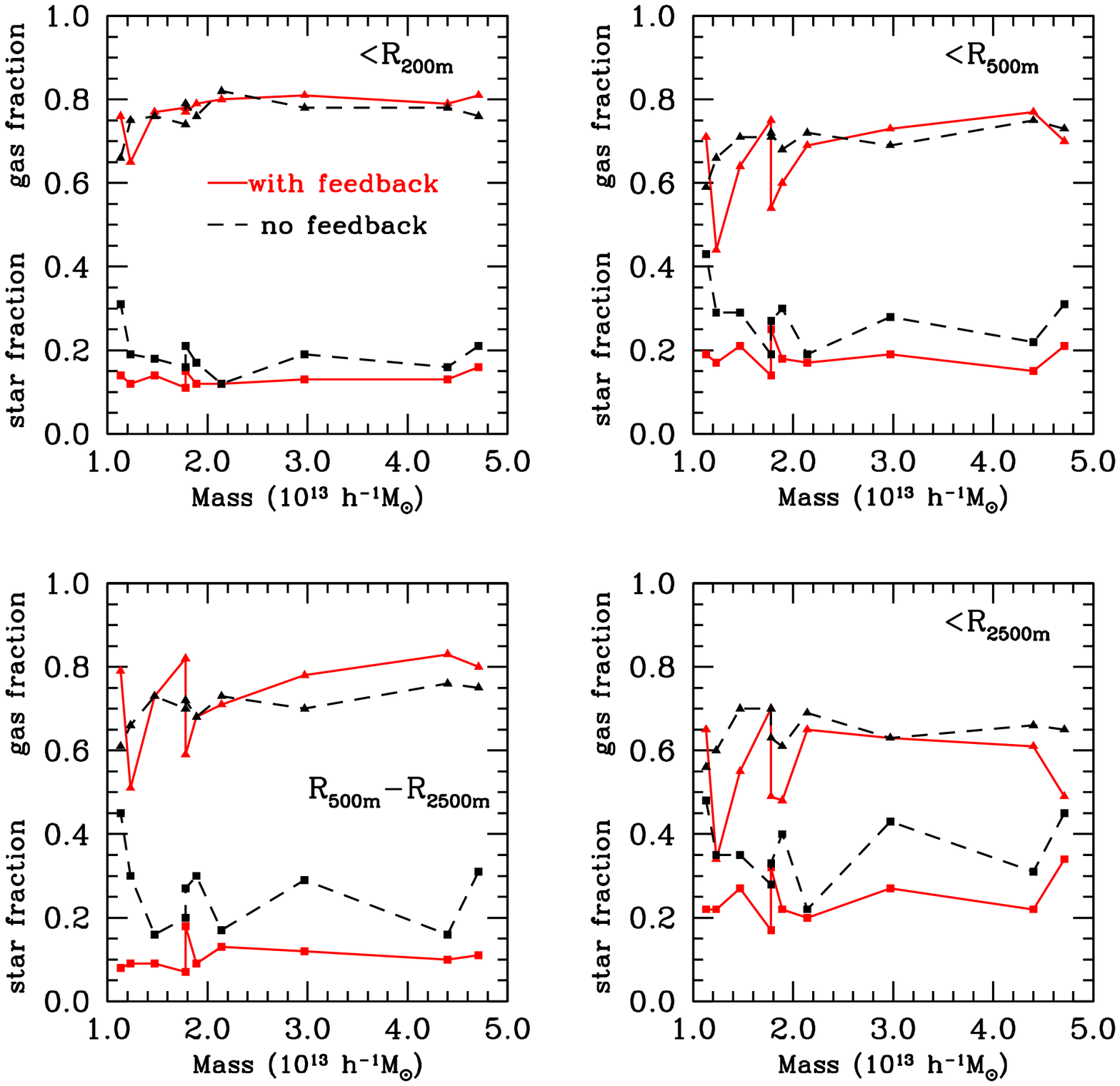}} 
    \caption{Cumulative gas and star fractions for the 10 most massive groups at $z=1$ measured within a radius $R=R_{200m}$ (top left), $R=R_{500m}$ (top right), and $R=R_{2500m}$ (lower right), and
    between $R=R_{500m}$ and $R=R_{2500m}$ (lower left). For each panel, squares represents the star fraction and triangles the gas fraction. Solid lines correspond to the simulation including quasar feedback and dotted lines represent the no-feedback case.}
    \label{baryons_vs_halomass}
  \end{center}
\end{figure*}

Quasar feedback has an equally significant effect on the gas distribution. As shown in the top panel of 
Fig.~\ref{baryon_profile}, hot gas is being driven out from the internal region of the halo ($R<R_{500m}$) towards the outer region. The gas density is lowered by 20-30\% in the core; to compensate for this depletion, gas density is 10\% higher at $R>0.3 R_{200m}$ compared to the no-feedback case. As is evident from the cumulative gas distribution, the feedback is not powerful enough to drive the gas from gravitational well of the halo. Note that there is still a difference in total gas mass of around 4\% within a radius of $2 R_{200m}$ which compensates for the lower star formation in these halos. 

Figure~\ref{baryons_vs_halomass} shows cumulative gas and star fractions as a function of halo mass, measured out to radii $R_{200m}$, $R_{500m}$, and $R_{2500m}$, and also between $R_{500m}$ and  $R_{2500m}$. Table~3 to 6 gives the fractions for individual halos at these radii and also the mean and scatter. On average, cumulative star fractions shows a 30\% depletion at all radii $<R_{500m}$ in simulation with quasar feedback; 
 Gas fractions shows only mild change at $R_{200m}$ and $R_{500m}$, although at $R_{2500m}$ the gas fraction is about 15\% lower 
with quasar feedback. When halo cores are excluded (i.e.\ between $R_{500m}$ and $R_{2500m}$), the gas fraction is enhanced by about 10\% in simulation with feedback. This again shows that gas is driven off from the inner region of the halos to outer region. The gas fraction $<R_{500m}$ displays a slight trend with mass in both simulations, although the star fraction shows no such effect. 

\begin{table}
\begin{center}
\begin{tabular}{ccccc}
\hline
\multicolumn{1}{c}{Mass} & \multicolumn{2}{c}{Quasar Feedback} & \multicolumn{2}{c}{No Quasar Feedback}\\
$10^{13}$ $M_{\odot}h^{-1}$  & $f_{gas}$ & $f_{star}$ & $f_{gas}$ & $f_{star}$ \\ \hline
4.71 & 0.81 & 0.16 & 0.76 & 0.21 \\
4.40 & 0.79 & 0.13 & 0.78 & 0.16 \\
2.97 & 0.81 & 0.13 & 0.78 & 0.19 \\
2.14 & 0.80 & 0.12 & 0.82 & 0.12 \\
1.89 & 0.79 & 0.12 & 0.76 & 0.17 \\
1.78 & 0.77 & 0.15 & 0.79 & 0.21 \\
1.78 & 0.78 & 0.11 & 0.74 & 0.16 \\
1.47 & 0.77 & 0.14 & 0.76 & 0.18 \\
1.23 & 0.65 & 0.12 & 0.75 & 0.19 \\
1.13 & 0.76 & 0.14 & 0.66 & 0.31 \\\hline
Mean & 0.77 & 0.13 & 0.76 & 0.19 \\
Scatter & 0.21 & 0.13 & 0.20 & 0.21\\ \hline
\end{tabular}
\end{center}
\caption{Cumulative fractions of gas and stars out to $R_{200m}$, both with and without quasar feedback.}
\label{baryonfrac_r200}
\end{table}

\begin{table}
\begin{center}
\begin{tabular}{ccccc}
\hline
\multicolumn{1}{c}{Mass} & \multicolumn{2}{c}{Quasar Feedback} & \multicolumn{2}{c}{No Quasar Feedback}\\
$10^{13}$ $M_{\odot}h^{-1}$ & $f_{gas}$ & $f_{star}$ & $f_{gas}$ & $f_{star}$ \\ \hline
4.71 & 0.70 & 0.21 & 0.73 & 0.31 \\
4.40 & 0.77 & 0.15 & 0.75 & 0.22 \\
2.97 & 0.73 & 0.19 & 0.69 & 0.28 \\
2.14 & 0.69 & 0.17 & 0.72 & 0.19 \\
1.89 & 0.60 & 0.18 & 0.68 & 0.30 \\
1.78 & 0.54 & 0.25 & 0.72 & 0.27 \\
1.78 & 0.75 & 0.14 & 0.71 & 0.19 \\
1.47 & 0.64 & 0.21 & 0.71 & 0.29 \\
1.23 & 0.44 & 0.17 & 0.66 & 0.29 \\
1.13 & 0.71 & 0.19 & 0.59 & 0.43 \\\hline
Mean & 0.66 & 0.18 & 0.70 & 0.28  \\
Scatter & 0.31 & 0.18 & 0.20 & 0.25 \\\hline
\end{tabular}
\end{center}
\caption{Same as in Table~\ref{baryonfrac_r200}, for $R_{500m}$.}
\label{baryonfrac_r500}
\end{table}

\begin{table}
\begin{center}
\begin{tabular}{ccccc}
\hline
\multicolumn{1}{c}{Mass} & \multicolumn{2}{c}{Quasar Feedback} & \multicolumn{2}{c}{No Quasar Feedback}\\
$10^{13}$ $M_{\odot}h^{-1}$ & $f_{gas}$ & $f_{star}$ & $f_{gas}$ & $f_{star}$ \\ \hline
4.71 & 0.49 & 0.34 & 0.65 & 0.45 \\
4.40 & 0.61 & 0.22 & 0.66 & 0.31 \\
2.97 & 0.63 & 0.27 & 0.63 & 0.43\\
2.14 &  0.65 & 0.20 & 0.69 & 0.22 \\
1.89 & 0.48 & 0.22 & 0.61 & 0.40 \\
1.78 & 0.49 & 0.32 & 0.63 & 0.33 \\
1.78 & 0.70 & 0.17 & 0.70 & 0.28 \\
1.47 &  0.55 & 0.27 & 0.70 & 0.35 \\
1.23 & 0.34 & 0.22 & 0.60 & 0.35 \\
1.13 &  0.65 & 0.22 & 0.56 & 0.48 \\ \hline
Mean & 0.56 & 0.24 & 0.64 & 0.36 \\ 
Scatter & 0.11 & 0.054 & 0.046 & 0.081 \\ \hline

\end{tabular}
\end{center}
\caption{Same as in Table~\ref{baryonfrac_r200}, for $R_{2500m}$.}
\label{baryonfrac_r2500}
\end{table}

\begin{table}
\begin{center}
\begin{tabular}{ccccc}
\hline
\multicolumn{1}{c}{Mass} & \multicolumn{2}{c}{Quasar Feedback} & \multicolumn{2}{c}{No Quasar Feedback}\\
$10^{13}$ $M_{\odot}h^{-1}$ & $f_{gas}$ & $f_{star}$ & $f_{gas}$ & $f_{star}$ \\ \hline
4.71 & 0.80 & 0.11 & 0.75 & 0.31 \\
4.40 & 0.83 & 0.10 & 0.76 & 0.16 \\
2.97 & 0.78 & 0.12 & 0.70 & 0.29 \\
2.14 & 0.71 & 0.13 & 0.73 & 0.17 \\
1.89 & 0.68 & 0.09 & 0.68 & 0.30 \\
1.78 & 0.59 & 0.18 & 0.72 & 0.27 \\
1.78 & 0.82 & 0.07 & 0.70 & 0.20 \\
1.47 & 0.73 & 0.09 & 0.73 & 0.16 \\
1.23 & 0.51 & 0.09 & 0.66 & 0.30 \\
1.13 & 0.79 & 0.08 & 0.61 & 0.45 \\\hline
Mean & 0.72 & 0.11 & 0.70 & 0.26 \\
Scatter & 0.10 & 0.032 & 0.04 & 0.09\\\hline
\end{tabular}
\end{center}
\caption{Cumulative fractions of gas and stars between radii $R_{500m}$ and $R_{2500m}$, both
with and without quasar feedback.}
\label{baryonfrac_outer}
\end{table}

\begin{figure}
  \begin{center}
    \begin{tabular}{cc}
      \resizebox{85mm}{!}{\includegraphics{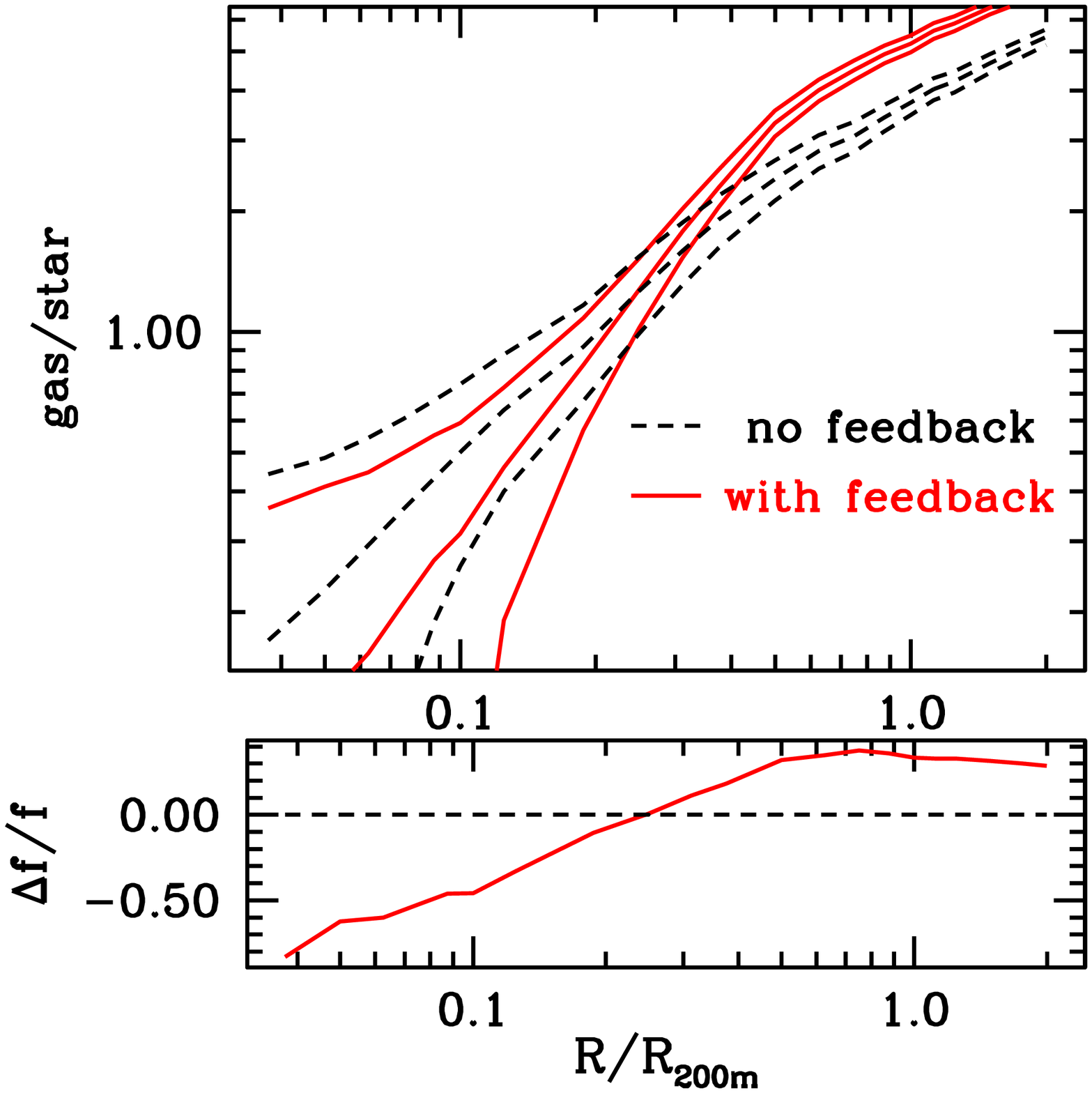}} \\
    \end{tabular}
    \caption{The average cumulative fraction of the ratio of gas mass and stellar  mass for the ten halos. Solid lines represent the mean and scatter for the sample including quasar feedback, while dotted lines represent the same for the no-feedback case.}
    \label{gas_star_ratio}
  \end{center}
\end{figure}

Figure~\ref{gas_star_ratio} shows the cumulative ratio of gas to stars. This quantity plays an important role for determining the cosmic matter density \citep{white93, evrard97, allen_etal02, ettori04, allen_etal04}. Usually it is assumed that this ratio is fixed at any radius with negligible redshift evolution \citep{ettori06}. we find that this assumption does not hold for either of the simulations. Without quasar feedback, the gas mass to stellar mass ratio changes roughly from 2 to 5, a factor of 2.5, between $0.3 R_{200m}$ and $R_{200m}$; for the simulation including quasar feedback the corresponding change in the ratio is slightly larger, from 2 to 7.5, a factor of 3.5. The ratio rises more steeply for the simulation with feedback and continues increasing beyond $R_{200m}$.


\subsection{Thermal Sunyaev-Zeldovich Decrements}

\begin{figure}
  \begin{center}
    \begin{tabular}{cc}
      \resizebox{85mm}{!}{\includegraphics{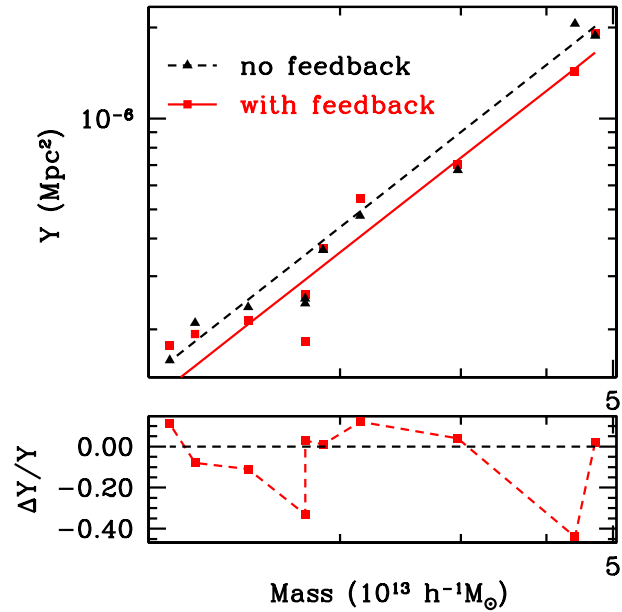}} 
    \end{tabular}
    \caption{Sunyaev-Zeldovich $Y$-distortion versus halo mass for 10 halos, for mass and gas
    within $1\,{\rm Mpc}/h$ of the halo center. Squares represent values from simulation D4 including quasar feedback; triangles represent values from simulation D4 without feedback. Lines are the best-fit power law to the $Y$-mass relation including quasar feedback (solid)  and without quasar feedback (dotted). The lower panel shows the fractional change in $Y$ between the two simulations.}
    \label{SZ_mass}
  \end{center}
\end{figure}

The thermal Sunyaev-Zeldovich distortion from quasar feedback has been studied previously (see, e.g., \cite{sk07, evan07} and references therein). This effect has a systematic impact on galaxy group-sized halos. As discussed above, the inaccuracy in the pressure profile due to numerical resolution limitations 
is on the order of 10\% for $R<0.1 R_{200m}$, so we exclude the halo core region when 
calculating SZ distortions. This does not substantially affect any of our results since the major contribution to 
the SZ signal comes from the region outside the halo cores \citep{ks02}. We calculate the mean
Compton $y$-distortion, which we denote $Y$, by integrating the gas pressure along the line of sight for each halo out to a radius of $R_{200m}$ and over the projected cross-section of the cluster in comoving coordinates. 
Figure~\ref{SZ_mass} shows the 
$Y$ versus mass for the halos considered here, both with and without quasar feedback, with the lower panel showing the fractional change in $Y$ . The individual halo
$Y$-parameters are given in Table~\ref{y_m_table}. On average, the $Y$ parameter
changes by 6\% (excluding the mergers) due to quasar feedback in these galaxy groups. Note that Y parameter in the run with the feedback shows both increase and decrease compared to the no feedback run as a function of mass. 

We also give a power law fit to the $Y$-mass relation of the form $Y/E(z)^{2/3}=10^{\beta} 
(M_{200m}/10^{14} M_\odot)^{\alpha}$ \citep{sehgal07}, where $\alpha$ and $\beta$ are fitting parameters and $E(z)=(\Omega_m(1+z)^3+\Omega_{\Lambda})^{0.5}$ is the redshift evolution of the Hubble parameter. Although the scatter is large, the power-law fits in both simulations (given in 
Table~\ref{y_m_powerlaws} are close, and the values are consistent with other studies with larger numbers of halos \citep{sehgal07}. 

As shown in \cite{ks02}, the SZ power spectrum receives a dominant contribution from high redshift halos; especially for $l>3000$, the contribution to $C_l$ comes mostly from $z>1$. The halo mass range considered here provides significant contribution to the $C_l$ for $l>5000$ and non-negligible contribution for $l=3000$ to 5000. Since $C_l \propto Y^2$, we expect that quasar feedback will
lead to a systematic increase in $C_l$ on the order of 10\% between $l=5000$ and 10000.

Note that the difference in $Y$ between the feedback and no-feedback cases does not tend to decrease with mass (Fig.~\ref{SZ_mass}), although the scatter is too large to claim any statistical significance of this behavior. It is imperative to 
simulate bigger volumes to quantify the effect of quasar feedback on the $Y$-mass relation for 
galaxy clusters,  and the corresponding systematic differences in cluster mass estimates. 
We  also emphasize that effect of quasar feedback generally increases with redshift, 
so our results at $z=1$ give conservative estimates of the quasar feedback impact on the SZ signal at earlier times.  
  
\begin{table}
\begin{center}
\begin{tabular}{cccc}
\hline
 $M_{200m}$ & $Y$ (feedback)&  $Y$ (no feedback)& $\Delta y/y$\\
 $10^{13} M_{\odot}/h$ & $10^{-7}\,{\rm Mpc}^2$ & $10^{-7}{\rm Mpc}^2$ & \%\\ \hline
4.71 & 1.91 & 1.88 & 0.02 \\
4.40 & 1.43 & 2.06 & -0.44\\
2.97 & 0.71 & 0.67 & 0.04 \\
2.14 & 0.54 & 0.48 & 0.12 \\
1.89 & 0.37 & 0.37 & 0.01 \\
1.78 & 0.26 & 0.25 & 0.03 \\
1.78 & 0.18 & 0.24 & -0.33\\
1.47 & 0.21 & 0.24 & -0.11 \\
1.23 & 0.19 & 0.21 & -0.08 \\
1.13 & 0.18 & 0.16 & 0.11 \\\hline
\end{tabular}
\end{center}
\caption{\rm The relation between SZ $Y$-distortion and cluster mass for galaxy groups 
with and without quasar feedback.}
\label{y_m_table}
\end{table}

\begin{table}
\begin{center}
\begin{tabular}{ccc}
\hline
 & $\alpha$ & $\beta$ \\ \hline
with feedback & 1.78 $\pm$ 0.06 & -5.55 $\pm$ 0.17\\
no feedback & 1.79 $\pm$ 0.05 & -5.47 $\pm$ 0.13\\\hline
\end{tabular}
\end{center}
\caption{\rm Power law fits to the SZ $Y$-mass relation for galaxy groups with and without quasar feedback, as displayed in Fig.~\ref{SZ_mass}.}
\label{y_m_powerlaws}
\end{table}

\section{Discussion} 

We have studied the effect of quasar feedback on baryon fractions and on thermodynamics of intracluster medium of intermediate mass halos corresponding to galaxy groups. Our analysis uses high-resolution N-body plus hydro cosmological simulations in a box with side length 33.75 Mpc/$h$. 
One simulation is conventional, while another incorporates black hole growth, accretion, and
energy ejection assuming simple astrophysics consistent with observations; both simulations use the
same initial conditions so individual large haloes can be compared. From the ten most massive
haloes, with masses ranging between $1$ and $5\times 10^{13} M_\odot/h$, we draw the
following conclusions:

1. Compared to the no-feedback case, star formation is suppressed by 30-40\% in the inner regions of the halos because of the additional pressure support provided by quasar feedback.

2. Quasar feedback redistributes hot gas, driving it from the inner region towards the outer part of the halos. As a result, gas density is 20\% less in the inner part and 10\% to 15\% greater in the outer region when compared to the simulation without feedback. However, the gas fraction in the two simulation differs by only 5\% to 10\%, and gas fractions tends to increase mildly with increasing halo mass.


3. The ratio of gas mass to stellar mass increases by a factor of 3.5 in the simulation including quasar feedback and a factor of 2.5 in the simulation without quasar feedback in the region $0.2 R_{200m}<R<0.5 R_{200m}$. This contradicts the common assumption that this ratio is constant at all radii.

4. Both temperature and entropy increase by 30\% to 50\% in the halo core region because of 
the additional thermal energy radiated by quasars.

5. Pressure decreases by 30\% in the inner region and increases by 15\% to 20\% at radii larger than 0.4 $R_{200m}$ due to the
increased gas density in this region. This leads to a change of about 6\% in the mean Sunyaev-Zeldovich
$Y$-distortion. The resulting SZ angular power spectrum will be larger
by around 10\% for $l>5000$. We find little dependence of the SZ enhancement with halo mass.

The effects of quasar feedback on the intracluster medium will be most evident in the group-sized
haloes considered here, with their relatively shallow gravitational potential wells. Observationally,
the most interesting haloes are larger in mass by a factor of ten, galaxy clusters: these are the haloes which are most readily detected via their SZ, X-ray, or optical signals. 
The gas fractions do not show any particular trend with increasing halo mass, and star fractions 
increase very weakly with mass over the halo mass range studied here. So it is reasonable to 
expect that the results of this work will hold for cluster-sized halos as well. Nevertheless, given the  substantial impact of quasar feedback on various properties of the intracluster medium which the current study suggests, it is imperative to study cluster-sized halos as well. This requires
larger-volume simulations, as the number density of clusters decreases with cluster mass. To this end, we are currently running a simulation of box size $50\,{\rm Mpc}/h$; results will be reported elsewhere.

This is the first attempt to study the impact of quasar feedback on the baryon fraction and thermodynamics of the intracluster medium in a cosmological hydrodynamic simulation. Both the gas and star fractions in our simulation are consistent with current observational limits \citep{allen_etal04, ettori99}. Note that we have studied only quasar feedback at redshifts greater than unity. However,
active galactic nuclei also inject energy into the ICM via a ``radio mode'' which is believed to be the dominant feedback mechanism at lower redshift \citep{sijacki05, sijacki07}. Thus our results should be treated as a conservative estimate of the total impact of AGN feedback for galaxy groups at low redshifts. 

Gas pressure in cosmological halos, particularly those with masses ranging from galaxy groups to galaxy clusters, determines the important thermal Sunyaev-Zeldovich signal which will soon be
measured with high precision. The gas fraction is important for connecting kinematic SZ signals of
cluster gas momentum with theoretical predictions about cluster velocity or total momentum. This
paper takes the first step towards quantifying the impact of quasars on these quantities, which
turns out to be significant but not dominating. Much work remains to be done, both through
larger simulations which contain many galaxy-cluster-sized haloes, and in enhancing the
realism of the quasar feedback models. We hope the results here plus the exciting observational
prospects in the near future will open the door to further advances in this area. 


\section*{Acknowledgments}
It is a pleasure to acknowledge many helpful discussions with Andrey Kravtsov, particularly for an informative guide to previous work on related problems. Inti Pelupessy contributed useful discussions and help with the simulation codes, and Suchetana Chatterjee provided simulation analysis code. Simulations were performed at the Pittsburgh Supercomputing Center. We thank Volker Springel for making the GADGET code available to the community, without which these simulations would not have been possible. The authors would also like to thank the anonymous referee for useful suggestions. This work has been supported by NSF grant AST-0408698 to the ACT project, and
by NSF grant AST-0546035. SB has been partly supported by a Mellon Graduate Fellowship
at the University of Pittsburgh.

\bibliographystyle{mn2e}

\bibliography{paper2final}

\end{document}